\documentclass[journal]{IEEEtran}
\usepackage{amsfonts}
\usepackage{amssymb}
\usepackage{amsthm}
\usepackage{amsmath,amsfonts,amssymb}
\usepackage[dvips]{graphicx}
\usepackage{verbatim}
\usepackage{setspace}
\usepackage{bm}
\usepackage{algorithmic} 
\usepackage[ruled,vlined]{algorithm2e}
\usepackage{cite}
\usepackage{changepage}

\usepackage[bookmarks=false]{}

\newcommand{\figwidth}{9}
\IEEEoverridecommandlockouts

\begin{document}

\title{Codebook Design for Millimeter-Wave Channel Estimation with Hybrid Precoding Structure}

\author{Zhenyu Xiao,~\IEEEmembership{Member,~IEEE,}
        Pengfei Xia,~\IEEEmembership{Senior Member,~IEEE,}
and Xiang-Gen Xia,~\IEEEmembership{Fellow,~IEEE}
\thanks{This work was partially supported by the National Natural Science Foundation of China (NSFC) under grant Nos. 61571025, 91338106, 91538204, and 61231013, National Basic Research Program of China under grant No.2011CB707000, and Foundation for Innovative Research Groups of the National Natural Science Foundation of China under grant No. 61221061.}
\thanks{Z. Xiao is with the School of
Electronic and Information Engineering, Beijing Key Laboratory for Network-based Cooperative Air Traffic Management, and Beijing Laboratory for General Aviation Technology, Beihang University, Beijing 100191, P.R. China.}
\thanks{P. Xia is with the School of Electronics and Information Engineering and the Key Laboratory of Embedded System and Service Computing, Tongji University, Shanghai, P.R. China.}
\thanks{X.-G. Xia is with the Department of Electrical and Computer Engineering, University of Delaware, Newark, DE 19716, USA.}

\thanks{Corresponding Author: Dr. Z. Xiao with Email: xiaozy@buaa.edu.cn.}
}
\markboth{IEEE TRANSACTIONS ON WIRELESS COMMUNICATIONS, VOL. X, NO.
x, xxx 20xx}{Shell \MakeLowercase{\textit{et al.}}}

\maketitle
\begin{abstract}
In this paper, we study hierarchical codebook design for channel estimation in millimeter-wave (mmWave) communications with a hybrid precoding structure. Due to the limited saturation power of mmWave power amplifier (PA), we take the per-antenna power constraint (PAPC) into consideration. We first propose a metric, i.e., generalized detection probability (GDP), to evaluate the quality of \emph{an arbitrary codeword}. This metric not only enables an optimization approach for mmWave codebook design, but also can be used to compare the performance of two different codewords/codebooks. To the best of our knowledge, GDP is the first metric particularly for mmWave codebook design for channel estimation. We then propose an approach to design a hierarchical codebook exploiting BeaM Widening with Multi-RF-chain Sub-array technique (BMW-MS). To obtain crucial parameters of BMW-MS, we provide two solutions, namely a low-complexity search (LCS) solution to optimize the GDP metric and a closed-form (CF) solution to pursue a flat beam pattern. Performance comparisons show that BMW-MS/LCS and BMW-MS/CF achieve very close performances, and they outperform the existing alternatives under the PAPC.
\end{abstract}


\section{Introduction}
\IEEEPARstart{M}{illimeter-wave} (mmWave) communication is a promising technology for next-generation wireless communication owing to its abundant frequency spectrum resource, which promises a much higher capacity than the existing wireless local area networks (WLANs) \cite{wang_2011_MMWCS,Park_2010_11ad,Xia_2011_60GHz_Tech,xia_2008_prac_ante_traning,xia_2008_multi_stage,xia2012system,yong2012system,Lakkis2008,wang_2009_beam_codebook} and the current cellular mobile communication \cite{khan_2011,alkhateeb2014mimo,han2015large,roh2014millimeter,sun2014mimo,niu2015survey,wang2014tens,wang2015multi}.
In order to bridge the link budget gap due to the extremely high path loss in mmWave band, beamforming with large antenna arrays are generally required in mmWave communications. Subject to the expensive radio-frequency (RF) chains, analog beamforming/combining structure is usually preferred, where all the antennas share a single RF chain and have constant-amplitude (CA) constraint on their weights \cite{Xia_2011_60GHz_Tech, xia_2008_prac_ante_traning,wang_2009_beam_codebook}. Meanwhile, a hybrid analog/digital precoding/combining structure was also proposed to realize multi-stream/multi-user transmission \cite{alkhateeb2014mimo,roh2014millimeter,sun2014mimo}, where a small number of RF chains are tied to a large antenna array.

Subject to the hardware constraint, i.e., the number of RF chains is far less than the number of antennas in general, the conventional multiple-input multiple-output (MIMO) channel estimation is basically infeasible in mmWave communications either due to high pilot overhead or high computational cost, and new channel estimation methods need to be tailored to mmWave systems \cite{alkhateeb2014channel}. For the hybrid precoding structure, as mmWave channel is generally sparse in the angle domain, different compressed sensing (CS) based channel estimation methods were proposed to estimate the steering angles of multipath components (MPCs) \cite{alkhateeb2014channel,alkhateeb2014mimo,alkhateeb2015compressed,peng2015enhanced,wang2015multi}.
For the analog beamforming structure, a switched beamforming approach was usually adopted \cite{wang_2009_beam_codebook,kokshoorn2015fast}, where the beam search space (at the transmitter and receiver side, respectively) is represented by a codebook containing multiple codewords, and the best transmit/receive beams are found by searching through their respective codebooks.


In practical mmWave channel estimation, a coarse sub-codebook may be defined with a small number of coarse sectors (or low-resolution beams) covering the intended angle range, while a fine sub-codebook may be defined with a large number of fine (or high-resolution) beams covering the same intended angle range, and that a coarse sector may have the same coverage as that of multiple fine beams together \cite{xia2012system,yong2012system,Lakkis2008,wang_2009_beam_codebook}. A divide-and-conquer search may then be carried out across the hierarchical codebook, by finding the best sector first on the low-resolution codebook level, and then finding the best beam on the high-resolution codebook level, while the best high-resolution beam is encapsulated in the best sector \cite{xia2012system,yong2012system,Lakkis2008,wang_2009_beam_codebook}. Such a hierarchical codebook structure and the associated multi-stage beam search have been adopted in many recent works \cite{alkhateeb2014channel,alkhateeb2014mimo,alkhateeb2015compressed,peng2015enhanced,wang2015multi,wang_2009_beam_codebook,kokshoorn2015fast}.

Performances of the search schemes, including the search time and detection rate of desired MPCs, are highly dependent on the codebook design. With an analog beamforming structure, \cite{wang_2009_beam_codebook} proposed to use wider beams to speed up beam search, but design approaches to broaden the beams were not studied. In \cite{he2015suboptimal}, a binary-tree structured hierarchical codebook was designed by using brute-force antenna deactivation (DEACT), where wider beams were generated by turning off part of the antennas. In \cite{xiao2016codebook}, a hierarchical codebook was also designed, where beam widening is achieved via sub-array technique. Although it was shown to outperform DEACT \cite{xiao2016codebook}, half of antennas may still need to be turned off for some codewords. In brief, to design a full codebook with an analog beamforming structure, antenna deactivation is basically needed, which not only reduces the total transmission power, but also requires an analog switch in each antenna branch, leading to additional cost and power consumption \cite{adabi2010mm}.

In contrast, a hybrid precoding structure with multiple RF chains (typically a few) cannot only enable multi-stream transmission, but also offer higher flexibility for codebook design; hence antenna deactivation and analog switches can be avoided. In \cite{alkhateeb2014channel}, the hybrid precoding structure was (maybe firstly) adopted to shape wider beams by exploiting the sparse reconstruction approach (SPARSE), but high-quality wide beams can be shaped only when the number of RF chains is large enough and deep sinks within the angle range appear otherwise. In addition, a phase-shifted discrete Fourier Transform (PS-DFT) method was also proposed in \cite{song2015multiRes}, where wider beams are shaped by steering multiple RF chains to adjacent equally spaced angles; thus a large number of RF chains are required to shape a very-wide codeword. Although these works \cite{alkhateeb2014channel,song2015multiRes} are theoretically feasible, they basically need a lot of RF chains for very-wide codewords, which may make them unappropriate for devices with only a few RF chains.

On the other hand, with multiple RF chains the output powers of the antennas may be significantly different from each other due to combining signals of multiple RF chains, and the power fluctuation is expected to be more severe when the number of RF chains is greater. Since in mmWave integrated circuits the saturation power of a PA is usually limited \cite{jin2007millimeter,zhao2012wideband}, the output power fluctuation may limit the total transmission power. In these parallel works \cite{alkhateeb2014channel,song2015multiRes}, the per-antenna power constraint (PAPC), caused by the limited saturation power of PA in each antenna branch, was not taken into account.

In this paper, we target at designing a codebook for mmWave channel estimation with a hybrid precoding structure (typically with a few RF chains), and we take the PAPC into account in the design. We first propose a metric, called generalized detection probability (GDP), to evaluate the quality of an arbitrary codeword. This metric not only enables a general optimization approach for mmWave codebook design, but also can be used to compare the performance of two different codewords/codebooks. To the best of our knowledge, GDP is the first metric particularly for mmWave codebook design. We then propose an approach to design a hierarchical codebook for the hybrid structure, where BeaM is Widened via Multi-RF-chain Sub-array technique (BMW-MS). To obtain crucial parameters of BMW-MS, we provide two solutions, namely a low-complexity search (LCS) solution to optimize the GDP metric and a closed-form (CF) solution to pursue a flat beam pattern. Performance comparisons show that BMW-MS/LCS and BMW-MS/CF achieve almost equivalent performances, and they (with only 2 RF chains) outperform the existing alternatives under the PAPC.

The rest of this paper is as follows. In Section II, the system and channel models are introduced. In Section III, the channel estimation method is proposed, and the problem of codebook design is formulated. In Section IV, the GDP metric is proposed. In Section V, the hierarchical codebook design is presented. In Section VI, performance evaluation is conducted. The conclusions are drawn lastly in Section VII.

Symbol Notations: $a$, $\mathbf{a}$, $\mathbf{A}$, and $\mathcal{A}$ denote a scalar variable, a vector, a matrix, and a set, respectively. $(\cdot)^{\rm{*}}$, $(\cdot)^{\rm{T}}$ and $(\cdot)^{\rm{H}}$ denote conjugate, transpose and conjugate transpose, respectively. $\mathbb{E}(\cdot)$ denotes expectation operation. $[\mathbf{a}]_i$ and $[\mathbf{A}]_{ij}$ denote the $i$-th entry of $\mathbf{a}$ and the $i$-row and $j$-column entry of $\mathbf{A}$, respectively. $[\mathbf{a}]_{i:j}$, $[\mathbf{A}]_{:,j}$, and $[\mathbf{A}]_{j,:}$ denote a vector with entries being the $i$-th to $j$-th entries of $[\mathbf{a}]$, the $j$-th column of $\mathbf{A}$ and the $j$-th row of $\mathbf{A}$, respectively. $|\cdot|$, $\|\cdot\|$ and $\|\cdot\|_\infty$ denote the absolute value, two-norm and $\infty$-norm respectively.

\section{System and Channel Models}
\begin{figure*}[ht]
\begin{center}
  \includegraphics[width=16 cm]{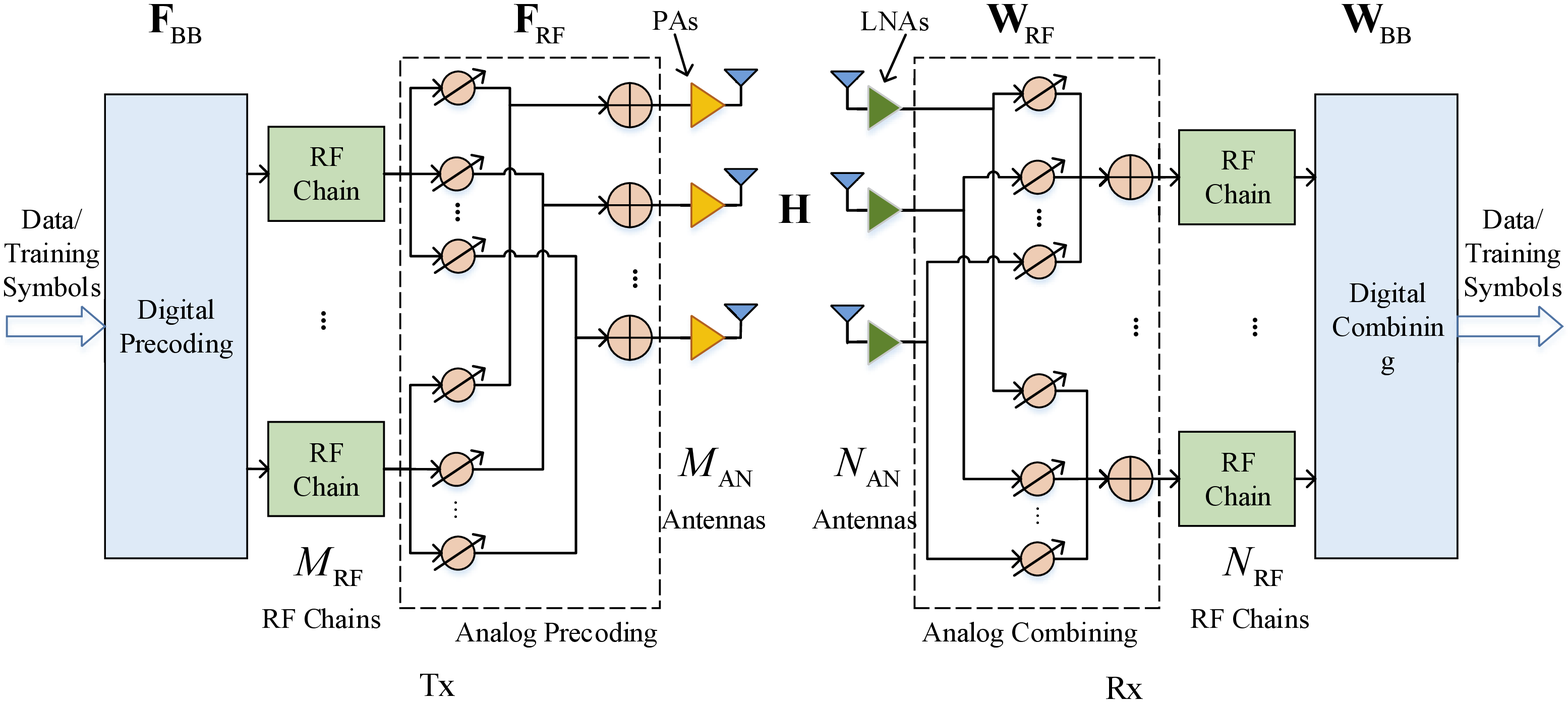}
  \caption{Illustration of a hybrid analog/digital precoding and combing structure with PAs.}
  \label{fig:system}
\end{center}
\end{figure*}

Without loss of generality, we consider a point-to-point mmWave system with a hybrid digital/analog precoding/combining structure, as shown in Fig. \ref{fig:system}, where multiple RF chains are tied to a half-wave spaced uniform linear array (ULA) at both the Tx and Rx. While extending the channel estimation method and the codebook design to a downlink multiuser scenario is straightforward. Relevant parameters are listed below, where $N_{\rm{ST}}$ is the number of data streams.
\begin{tabbing}
 ~~~\= $M_{\rm{RF}}$~~~~\=  The number of RF chains at the Tx.\\
 \> $M_{\rm{AN}}$ \>  The number of antennas at the Tx.\\
 \> $N_{\rm{RF}}$ \>  The number of RF chains at the Rx.\\
 \> $N_{\rm{AN}}$ \>  The number of antennas at the Rx.\\
 \> ${{\bf{F}}_{\rm{BB}}}$ \>  ${{M_{\rm{RF}}} \times {N_{\rm{ST}}}}$ digital precoding matrix at the Tx.\\
 \> ${{{\bf{F}}_{\rm{RF}}}}$ \>  ${{M_{\rm{AN}}} \times {M_{\rm{RF}}}}$ analog precoding matrix at the Tx.\\
 \> ${{\bf{W}}_{\rm{BB}}}$ \>  ${{N_{\rm{RF}}} \times {N_{\rm{ST}}}}$ digital combining matrix at the Rx.\\
 \> ${{\bf{W}}_{\rm{RF}}}$ \>  ${{N_{\rm{AN}}} \times {N_{\rm{RF}}}}$ analog combining matrix at the Rx.\\
 \> $\underline{{\bf{F}}}$ \>  A Tx \emph{composite codeword}, $\underline{{\bf{F}}}\triangleq ({{\bf{F}}}_{\rm{RF}},{{\bf{F}}}_{\rm{BB}})$.\\
 \> $\underline{{\bf{f}}}_{i}$ \>  A Tx codeword, $\underline{{\bf{f}}}_{i}\triangleq ({{\bf{F}}}_{\rm{RF}},[{{\bf{F}}}_{\rm{BB}}]_{:,i})$.\\
 \> $\underline{{\bf{W}}}$ \>  A Rx \emph{composite codeword}, $\underline{{\bf{W}}}\triangleq ({{\bf{W}}}_{\rm{RF}},{{\bf{W}}}_{\rm{BB}})$.\\
 \> $\underline{{\bf{w}}}_{i}$ \>  A Tx codeword, $\underline{{\bf{w}}}_{i}\triangleq ({{\bf{W}}}_{\rm{RF}},[{{\bf{W}}}_{\rm{BB}}]_{:,i})$.\\
  \end{tabbing}  \vspace{-0.15 in}
Basically, we have $ {M_{\rm{RF}}} \le {M_{\rm{AN}}}$ and ${N_{\rm{RF}}} \le {N_{\rm{AN}}}$, but in practical mmWave systems, ${M_{\rm{RF}}}$ and ${N_{\rm{RF}}}$ are small, basically far less than ${M_{\rm{AN}}}$ and ${N_{\rm{AN}}}$, respectively.

In this paper, we propose a channel estimation method based on a hierarchical codebook, and design the codebook with the hybrid structure in Fig. \ref{fig:system}. A Tx codebook is a collection of \emph{composite codewords}, and a Tx composite codeword is in fact a precoding matrix pair $({{{\bf{F}}_{\rm{RF}}}},~{{\bf{F}}_{\rm{BB}}})$, which can be seen as the composite of $M_{\rm{RF}}$ Tx codewords $\{({{{\bf{F}}_{\rm{RF}}}},~{{\bf{F}}_{\rm{BB}}}_{:,i})\}_{i=1,2,...,M_{\rm{RF}}}$. The constitution of the Rx codebook is similar to the Tx codebook. We emphasize that in this paper \emph{we use underline to mark a codeword and a composite codeword, respectively}, as shown in the above list. Note that the Tx/Rx codebooks are predesigned, and thus they are irrelevant to an instantaneous channel response. However, they are designed based on the steering feature of mmWave channel, and they are used to reduce the training overhead in channel estimation.

We emphasize that there is a single PA in each antenna branch right before the antenna at the Tx. Since in practice the saturation power of a mmWave PA is basically limited \cite{jin2007millimeter,zhao2012wideband}, we have the PAPC in our model, i.e., the saturation power of a PA is $P_{\rm{PEP}}$, which was not considered in \cite{alkhateeb2014channel} and \cite{song2015multiRes}. In addition, similar to \cite{alkhateeb2014channel} and \cite{song2015multiRes}, the analog precoding/combining matrices are with constant-amplitude (CA) elements, because they are controlled by phase shifters. Note that there is no switch in each antenna branch.



Without loss of generality, we adopt the same channel model as that in \cite{he2015suboptimal,alkhateeb2014channel,Ayach2014,hur2013millimeter,nsenga_2009,xiao2016codebook}, which is given by
\begin{equation} \label{eq_Channel}
{\bf{H}} = \sqrt {{{M_{\rm{AN}}}}{{N_{\rm{AN}}}}} \sum\limits_{\ell  = 1}^L {{\lambda _\ell }{\bf{a}}({{N_{\rm{AN}}}},{\Omega _\ell }){\bf{a}}{{({{M_{\rm{AN}}}},{\psi _\ell })}^{\rm{H}}}},
\end{equation}
where $\lambda_\ell$ is the complex coefficient of the $\ell$-th path, $L$ is the number of MPCs, ${\bf{a}}(\cdot)$ is the steering vector function, ${\Omega _\ell }$ and ${\psi _\ell }$ are cos(AoD) and cos(AoA) of the $\ell$-th path, respectively. Let ${\theta _\ell }$ and ${\varphi _\ell }$ denote the physical AoD and AoA of the $\ell$-th path, respectively; then we have ${\Omega _\ell } = \cos ({\theta _\ell })$ and ${\psi _\ell } = \cos ({\varphi _\ell })$. Therefore, ${\Omega _\ell }$ and ${\psi _\ell }$ are within the range $[-1~1]$. \emph{For convenience, in the rest of this paper the cosine angles ${\Omega _\ell }$ and ${\psi _\ell }$ are called AoDs and AoAs, respectively. Without particular statement, the angle domain implicitly means cosine angle domain.} Similar to \cite{xiao2015Iterative,alkhateeb2014channel}, $\lambda_\ell$ can be modeled to be complex Gaussian distributed, while ${\Omega _\ell }$ and ${\psi _\ell }$ can be modeled to be uniformly distributed within $[-1,1]$. ${\bf{a}}(\cdot)$ is a function of the number of antennas and AoD/AoA, and can be expressed as
\begin{equation}
\begin{aligned}
{\bf{a}}(N,\Omega ) =\frac{1}{{\sqrt N }}[e^ {j\pi 0\Omega},~e^{ j\pi 1\Omega },...,e^{j\pi (N - 1)\Omega}]^{\rm{T}},
\end{aligned}
\end{equation}
where $N$ is the number of antennas ($N$ is ${M_{\rm{AN}}}$ at the transmitter and $M_{\rm{R}}$ at the receiver), $\Omega$ is AoD or AoA. It is easy to find that ${\bf{a}}(N,\Omega )$ is a periodical function which satisfies ${\bf{a}}(N,\Omega )={\bf{a}}(N,\Omega +2)$. The channel matrix ${\bf{H}}$ also has power normalization $
\sum_{\ell=1}^{L}\mathbb{E}(|\lambda_\ell|^2)=1.$

\section{Channel Estimation and the Problem of Codebook Design}
\subsection{Channel Estimation}
Subject to the hardware constraint, i.e., the number of RF chains is far less than that of the antennas, mmWave channel estimation is generally to search the AoDs/AoAs of several strong MPCs one by one via \emph{beam search} in the angle domain \cite{alkhateeb2014channel,alkhateeb2015compressed,peng2015enhanced,wang2015multi,wang_2009_beam_codebook,kokshoorn2015fast}. In this subsection, we propose an improved beam search method to search one MPC with the hybrid precoding structure in Fig. \ref{fig:system}.

In order to estimate the AoD/AoA of an MPC, signal measurements must be carried out based on transmission of training sequences. In each measurement, multi-stream orthogonal training sequences are transmitted from Tx to Rx, with precoding matrices selected from a Tx codebook and combining matrices selected from a Rx codebook, respectively. Hence, we have the following signal model for a measurement:
\begin{equation} \label{eq_received_symb}
\begin{aligned}
{\bf{Y}} &= \sqrt P {{\bf{W}}}_{\rm{BB}}^{\rm{H}}{{\bf{W}}}_{\rm{RF}}^{\rm{H}}{\bf{H}}{{{\bf{F}}}_{\rm{RF}}}{{{\bf{F}}}_{\rm{BB}}}{\bf{S}} + {{\bf{W}}}_{\rm{BB}}^{\rm{H}}{{\bf{W}}}_{\rm{RF}}^{\rm{H}}{\bf{Z}},
\end{aligned}
\end{equation}
where $P$ is the transmission power per stream, ${\bf{H}}$ is the channel matrix, $\bf{Z}$ is a white Gaussian noise matrix with average power $N_0$, $[{\bf{S}}]_{j,:}$ is the $j$-th transmitted training sequence with a length of ${L_{\rm{S}}}$ at Tx, $[{\bf{Y}}]_{i,:}$ is the received sequence at the $i$-th RF chain at Rx, \emph{$({{{\bf{F}}}_{\rm{RF}}},{{{\bf{F}}}_{\rm{BB}}})$ and $({{\bf{W}}}_{\rm{RF}},{{\bf{W}}}_{\rm{BB}})$ are a selected Tx composite codeword and a selected Rx composite codeword, respectively.} We have $[{\bf{S}}]_{m,:}[{\bf{S}}]_{n,:}^{\rm{H}}=0$ when $i\neq j$ and $[{\bf{S}}]_{m,:}[{\bf{S}}]_{m,:}^{\rm{H}}={L_{\rm{S}}}$. Let us omit the noise for simplicity. Then we have
\begin{equation}
[{\bf{Y}}]_{i,:} = \sqrt P [{{\bf{W}}}_{\rm{BB}}]_{:,i}^{\rm{H}}{{\bf{W}}}_{\rm{RF}}^{\rm{H}}{\bf{H}}\sum_{m=1}^{M_{\rm{RF}}}{{{\bf{F}}}_{\rm{RF}}}[{{{\bf{F}}}_{\rm{BB}}}]_{:,m}[{\bf{S}} ]_{m,:}
\end{equation}
Thus,
\begin{equation} \label{eq_rho_ij}
\begin{aligned}
\rho_{i,j}&=[{\bf{Y}}]_{i,:}[{\bf{S}} ]_{j,:}^{\rm{H}} \\
&={L_{\rm{S}}} \sqrt P [{{\bf{W}}}_{\rm{BB}}]_{:,i}^{\rm{H}}{{\bf{W}}}_{\rm{RF}}^{\rm{H}}{\bf{H}}{{{\bf{F}}}_{\rm{RF}}}[{{{\bf{F}}}_{\rm{BB}}}]_{:,j}\\
&\triangleq {L_{\rm{S}}} \sqrt P{{\bf{w}}}_i^{\rm{H}}{\bf{H}}{\bf{f}}_{j},
\end{aligned}
\end{equation}
where $i=1,2,...,N_{\rm{RF}}$, and $j=1,2,...,M_{\rm{RF}}$, ${\bf{w}}_{i}={{\bf{W}}}_{\rm{RF}}[{{\bf{W}}}_{\rm{BB}}]_{:,i}$ and ${\bf{f}}_{j}={{\bf{F}}}_{\rm{RF}}[{{\bf{F}}}_{\rm{BB}}]_{:,j}$ correspond to the $i$-th Rx codeword $\underline{{\bf{w}}}_{i}\triangleq ({{\bf{W}}}_{\rm{RF}},[{{\bf{W}}}_{\rm{BB}}]_{:,i})$ of the selected Rx composite codeword and the $j$-th Tx codeword $\underline{{\bf{f}}}_{j}\triangleq ({{\bf{F}}}_{\rm{RF}},[{{\bf{F}}}_{\rm{BB}}]_{:,j})$ of the selected Tx composite codeword, respectively.


Afterwards, Rx obtains the optimal Tx/Rx codeword pair as
\begin{equation} \label{eq_opt_index}
(j^\star,i^\star) = \mathop {\arg \max}\limits_{(j,i)}~|\rho_{i,j}|^2,
\end{equation}
and feeds back $j^\star$ to Tx. Hence, in each measurement, Rx in fact finds the best Tx/Rx codeword pair with the highest signal power. If the Tx/Rx codewords are pre-designed to cover different angle ranges, then the AoD/AoA of the MPC will be within the angle coverage of the best Tx/Rx codewords, respectively.

\begin{figure}[t]
\begin{center}
  \includegraphics[width=8 cm]{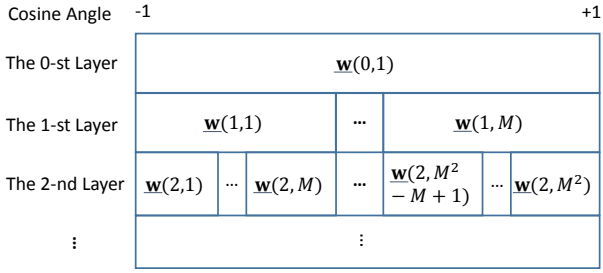}
  \caption{Beam coverage of a hierarchical codebook.}
  \label{fig:codebook}
\end{center}
\end{figure}

To reduce the training overhead, a hierarchical Tx or Rx codebook, which is a collection of codewords $\underline{{\bf{f}}}$ or $\underline{{\bf{w}}}$, is defined as Fig. \ref{fig:codebook}, The codebook has $\log_M(N)+1$ layers with indices from $k=0$ to $k=\log_M(N)$, where $M$ and $N$ are the number of RF chains and antennas, respectively. The number of codewords in the $k$-th layer is $M^k$, and
\begin{equation} \label{eq_beamcover_w}
\begin{aligned}
{\cal C}{\cal V}(\underline{{\bf{w}}}(k,n)) = [ - 1 + \frac{{2n - 2}}{{{M^k}}}, - 1 + \frac{{2n}}{{{M^k}}}],\\
k=0,1,...,\log_M(N),~n = 1,2,...,{M^k},
\end{aligned}
\end{equation}
where ${\cal C}{\cal V}(\underline{{\bf{w}}})$ denotes the beam coverage in the angle domain of codeword $\underline{{\bf{w}}}$, $\underline{{\bf{w}}}(k,n)$ denotes the $n$-th codeword in the $k$-th layer. Note that a set of $M$ adjacent codewords in the same layer, i.e., $\{\underline{{\bf{w}}}(k,(i-1)M+j)\}_{j=1,2,...,M}$ ($i=1,...,M^{k-1}$), constitute a composite codeword $\underline{{\bf{W}}}(k,i)$.


Based on a hierarchical codebook, an efficient divide-and-conquer search, as shown in Algorithm \ref{alg:search}, is launched to fast estimate the response of an MPC. It is noteworthy that a codebook shown in Fig. \ref{fig:codebook} and designed in this paper cannot only used in Algorithm \ref{alg:search}, but also in other beam search methods. On the other hand, only one MPC is searched out by launching Algorithm \ref{alg:search} once. For the case that multiple MPCs are needed to be searched out, Algorithm \ref{alg:search} must be launched multiple times to estimate different MPCs, one MPC at a time with new training sequences transmitted. However, extending the one-MPC search to multiple-MPC search is not trivial, because the contribution of the already searched MPC must be subtracted when searching a new MPC. Details may refer to \cite{alkhateeb2014channel,xiao2016lowcomplexity}.

\begin{algorithm}[t]\caption{Beam Search Algorithm Based on a Hierarchical Codebook with a Hybrid Structure.}\label{alg:search}
\textbf{1) Initialization:}\\
$k=1$. $/*$The layer index.$*/$\\
$k_{\rm{M}}=\max\{\log_{M_{\rm{RF}}}(M_{\rm{AN}}),\log_{N_{\rm{RF}}}(N_{\rm{AN}})\}$. $/*$The maximal layer.$*/$\\
Predefine Tx codebook $\mathcal{F}$ and Rx codebook $\mathcal{W}$.\\
$j_{\rm{T}}=i_{\rm{R}}=1$. $/*$Indices of Tx/Rx composite codewords.$*/$\\

\vspace{0.1 in}
\textbf{2) Iteration:}\\
\For{$k =1: k_{\rm{M}}$}
{
    Tx/Rx, respectively, selects a composite codeword $\underline{{\bf{F}}}(k,j_{\rm{T}})$/$\underline{{\bf{W}}}(k,i_{\rm{R}})$ from $\mathcal{F}$/$\mathcal{W}$, and sets the matrices in the composite codewords to the corresponding precoding/combining matrices.

    \vspace{0.05 in}

    Tx sends parallel orthogonal sequences ${\bf{S}}$ as \eqref{eq_received_symb}, and Rx receives and computes $\rho_{i,j}$ as \eqref{eq_rho_ij}.

    \vspace{0.05 in}
    Rx computes the optimal Tx/Rx index pair $(j^\star,i^\star)$ as \eqref{eq_opt_index}, and sets $i_{\rm{R}}=N_{\rm{RF}}*(i_{\rm{R}}-1)+i^\star$. Rx feeds back $j^\star$ to Tx, and Tx sets $j_{\rm{T}}=M_{\rm{RF}}*(j_{\rm{T}}-1)+j^\star$.
}

\vspace{0.1 in}
\textbf{3) Result:}

    The response of the estimated MPC is given by ${\bf{H}}_1=\rho_{i^\star, j^\star}{\bf{a}}({{N_{\rm{AN}}}},- 1 + \frac{{2i_{\rm{R}} - 1}}{{N_{\rm{AN}}}}){\bf{a}}{{({{M_{\rm{AN}}}},- 1 + \frac{{2j_{\rm{T}} - 1}}{{M_{\rm{AN}}}})}^{\rm{H}}}$.
\end{algorithm}

Another comment we want to make here is as follows. There is another dimension reduced channel estimation in massive MIMO system. It is to add a digital dimension reduced precoding matrix to feed Tx antennas \cite{larsson2014massive,you2015pilot}. However, this approach needs to decide which precoding matrix to use, which is, in fact, the main issue to address here.

Let us next evaluate the training overhead of Algorithm \ref{alg:search}. We adopt the duration of a training symbol as a unit to count the training overhead, which is the same as that in regular MIMO systems \cite{larsson2014massive,you2015pilot}. To guarantee the orthogonality between different training sequences, we have $L_{\rm{S}}\geq M_{\rm{RF}}$. It is noted that in regular MIMO systems $L_{\rm{S}}$ is usually set equal to $M_{\rm{RF}}$ to save overhead \cite{larsson2014massive,you2015pilot}, but in mmWave communications $L_{\rm{S}}$ is usually set much greater than $M_{\rm{RF}}$ to provide spreading gain for channel estimation, because Tx/Rx array gains are not yet available before channel estimation \cite{xiao2016codebook,alkhateeb2014channel}.

For a fair comparison with the alternatives, i.e., \cite{xiao2016codebook} and \cite{alkhateeb2014channel}, we assume that the length of the training sequence is the same ($L_{\rm{S}}$) in these algorithms.
In each measurement of Algorithm \ref{alg:search}, $M_{\rm{AN}}$ training sequences $\{[{\bf{S}}]_{j,:}\}_{j=1,2,...,M_{\rm{AN}}}$ are transmitted in parallel; so the training overhead is $L_{\rm{S}}$. Suppose ${M_{\rm{RF}}}={N_{\rm{RF}}}$ and $M_{\rm{AN}}=N_{\rm{AN}}$. The total training overhead of Algorithm \ref{alg:search} is $L_{\rm{S}}\log_{M_{\rm{RF}}}(M_{\rm{AN}})$, significantly less than that of the beam search algorithm with an analog beamforming structure in \cite{xiao2016codebook} ($2L_{\rm{S}}M_{\rm{RF}}\log_{M_{\rm{RF}}}(M_{\rm{AN}})$), as well as that of the beam search algorithm with a hybrid precoding structure in \cite{alkhateeb2014channel} (Algorithm 1 therein, which needs $L_{\rm{S}}M_{\rm{RF}}\log_{M_{\rm{RF}}}(M_{\rm{AN}})$ units if $K=M_{\rm{RF}}$). In brief, Algorithm \ref{alg:search} reduces the training overhead by a factor of the number of RF chains compared with the alternatives, which benefits from the parallel transmission of multiple training sequences. However, the cost of this benefit is that the codewords within a composite codeword must share the same analog precoding matrix, which will be considered in the codebook design.

\subsection{The Problem of Codebook Design}
As we can see, a codebook is critical to Algorithm \ref{alg:search}. In this paper we want to design a hierarchical codebook with the beam coverage shown in Fig. \ref{fig:codebook} based on the hybrid structure shown in Fig. \ref{fig:system}. We emphasize that we have both the PAPC on PAs and the CA constraint on the analog precoding/combining matrices. Since a Rx codebook design can be the same as a Tx codebook design, we proceed with Tx codebook design.


According to \eqref{eq_rho_ij}, with the hybrid structure an arbitrary codeword $\underline{{\bf{w}}}\triangleq ({{\bf{F}}_{\rm{RF}}},{{\bf{F}}_{\rm{BB}}}_{:,i})$ shapes an antenna weight vector (AWV) ${\bf{w}}={{\bf{F}}_{\rm{RF}}}{{\bf{F}}_{\rm{BB}}}_{:,i}$, and the beam steering and coverage of $\underline{{\bf{w}}}$ are in fact reflected by ${\bf{w}}$. Hence, codebook design in this paper is to design $\underline{{\bf{w}}}(k,n)$ such that ${\bf{w}}(k,n)$ has the beam coverage ${\cal C}{\cal V}({{\bf{w}}}(k,n))={\cal C}{\cal V}(\underline{{\bf{w}}}(k,n))$. \emph{For convenience, we also call ${\bf{w}}$ a codeword in the remaining of this paper,} but we emphasize that we want to design $\underline{{\bf{w}}}\triangleq ({{\bf{F}}_{\rm{RF}}},{{\bf{F}}_{\rm{BB}}}_{:,i})$ rather than just ${\bf{w}}$ itself, because ${\bf{w}}$ is solely determined by $\underline{{\bf{w}}}$ but not vice versa.

Consequently, a codeword ${\bf{w}}$ has the following structure:
\begin{equation}\label{eq_codeword_Tx}
{\bf{w}}={{\bf{F}}_{\rm{RF}}}{{\bf{f}}_{\rm{BB}}}=\sum\limits_{j = 1}^{{M_{\rm{RF}}}} [{{\bf{f}}_{\rm{BB}}}]_j[{{\bf{F}}_{\rm{RF}}}]_{:,j},
\end{equation}
where ${{\bf{f}}_{\rm{BB}}}={{\bf{F}}_{\rm{BB}}}_{:,i}$, $|[{{\bf{F}}_{\rm{RF}}}]_{:,j}|=\frac{1}{\sqrt{M_{\rm{AN}}}}{\bf{1}}$ (the CA constraint). Note that the codewords belong to the same composite codeword share the same ${{\bf{F}}_{\rm{RF}}}$, which must be considered in the design.

Given the target beam pattern of ${\bf{w}}(k,n)$ shown in Fig. \ref{fig:codebook}, we need to design
$({{\bf{F}}_{\rm{RF}}},~{{\bf{f}}_{\rm{BB}}})$ for each ${\bf{w}}(k,n)$, which is challenging due to the CA constraint on ${{\bf{F}}_{\rm{RF}}}$. In \cite{alkhateeb2014channel}, this problem is solved by exploiting the sparse reconstruction approach (SPARSE). While in \cite{song2015multiRes}, the problem is further constrained by letting $|{{\bf{f}}_{\rm{BB}}}={\bf{1}}|$, i.e., the transmission power of each RF chain is the same, i.e., $|[{{\bf{f}}_{\rm{BB}}}]_j|^2\|[{{\bf{F}}_{\rm{RF}}}]_{:,j}\|^2=1$. In such a case, a codeword is a combination of multiple RF vectors with equal power, and it is intuitive that by steering these RF vectors to equally spaced angles, a wide beam can be shaped. This is just the PS-DFT codebook proposed in \cite{song2015multiRes}.

In this paper, we also let $|{{\bf{f}}_{\rm{BB}}}|={\bf{1}}$ to simplify the problem, just the same as \cite{song2015multiRes}. However, we will propose different methods to design the codewords. In the following, we will first establish a general metric, i.e., the GDP metric, to evaluate the quality of an arbitrary Tx codeword ${\bf{w}}$. We emphasize that the metric is applicable for codewords with both analog and hybrid structures. Then we will design a hierarchical codebook with the target beam coverage shown in Fig. \ref{fig:codebook} with the codeword structure \eqref{eq_codeword_Tx}.

\section{The GDP Metric}
Given an arbitrary target codeword to cover an angle range $[\psi_0,\psi_0+B]$, there are many approaches to design it. It is clear that the best codeword should have constant absolute beam gain within the covered angle range (i.e., a totally flat beam pattern) \cite{song2015multiRes}. However, due to the CA constraint on the analog precoding/combining matrices, an ideal codeword can be hardly designed. Hence, suboptimal designs are of interest, and there have been many approaches to design a hierarchical codebook\cite{song2015multiRes,alkhateeb2014channel}. To the best of our knowledge, however, there is no particular metric to directly evaluate the quality of a codeword in the regime of mmWave communications. We can only judge the quality of a codeword by numerical simulation. Hence, in this subsection, we establish a general metric and introduce its properties and significance.

\subsection{The GDP Metric}
Let us first define the beam gain of an arbitrary codeword ${\bf{w}}$ along angle $\Omega$ ($\Omega\in[-1,1]$), i.e., $A({\bf{w}},\Omega )$:
\begin{equation} \label{eq_beam_gain_Define}
A({\bf{w}},\Omega ) = \sqrt N {\bf{a}}{(N,\Omega )^{\rm{H}}}{\bf{w}} = \sum\limits_{n = 1}^N {{{[{\bf{w}}]}_n}{e^{ - j\pi (n - 1)\Omega }}}.
\end{equation}

Intuitively, good codewords should have flat beam patterns, and mean square error (MSE) can be adopted to measure how flat a beam pattern is. Moreover, in mmWave communications, the saturation power of PA is limited. Hence, good codewords should also allow as high as possible maximal transmission power (MTP), which is limited by the saturation power of PA in each antenna branch. For instance, DEACT is not with high quality, because a lot of antenna elements are turned off, which significantly lowers the MTP. In fact, under the PAPC, the MTP of an arbitrary codeword ${\bf{w}}$ is given by
\begin{equation} \label{eq_MTP}
P_{\rm{MAX}}({\bf{w}})=\frac{P_{\rm{PER}}}{\max(\{|[{\bf{w}}]_n|^2\}_{n=1}^{N})}\triangleq \frac{P_{\rm{PER}}}{\|{\bf{w}}\|_\infty^2},
\end{equation}
where $P_{\rm{PER}}$ is the PAPC in this paper, i.e., the saturation power of PA in each antenna branch. It is clear that given fixed $P_{\rm{PER}}$, $P_{\rm{MAX}}({\bf{w}})$ is maximized when ${\bf{w}}$ (with unit 2-norm) has CA elements.

As we can see, both MSE and MTP may affect the quality of a codeword and their effects are different. In fact, these two metrics are basically contradictory to each other, i.e., a codeword may have a small MSE but meanwhile also a small MTP. In general, we want small MSE along with large MTP. It is not favorable to define a metric with simple operations between the MSE and MTP. For rigorousness, we directly bridge the metric to the detection performance in beamforming training, because the codebook is particularly designed for it. During beamforming training, many Tx/Rx codeword pairs will be selected to detect the AoD/AoA of an MPC. When the AoD/AoA of the MPC locate within the coverage of the codewords, the detection probability (DP) is a direct and exact metric. Hence, we can derive the average DP, and generalize a metric based on the average DP for the Tx codewords.

Suppose that Tx transmits a training sequence with codeword ${\bf{w}}_{\rm{T}}$, and Rx receives with codeword ${\bf{w}}_{\rm{R}}$, i.e., ${\bf{w}}_{\rm{T}}$ and ${\bf{w}}_{\rm{R}}$ are fixed. The target beam coverage of ${\bf{w}}_{\rm{T}}$ is $[\psi_0,\psi_0+B]$. We want to develop a metric to evaluate the Tx codeword ${\bf{w}}_{\rm{T}}$ based on the average DP of a single MPC.

Let ${\bf{H}}_0$ denote the channel response for the MPC to be detected, and it can be defined as
\begin{equation}
{\bf{H}}_0 = \sqrt {{{M_{\rm{AN}}}}{{N_{\rm{AN}}}}} \lambda {\bf{a}}(N_{\rm{AN}},\Omega){\bf{a}}(M_{\rm{AN}},\psi)^{\rm{H}},
\end{equation}
where $\lambda$, $\Omega$ and $\psi$ denote the gain, AoA and AoD of the MPC, respectively. Without loss of generality, we assume $\lambda\thicksim \mathcal{CN}(0,1)$, $\Omega$ and $\psi$ are uniformly distributed within $[-1,1]$.

Given ${\bf{H}}_0$ the detection problem can be formulated as binary hypothesis testing given by \cite{song2015multiRes}
\begin{equation} \label{eq_DetectionProb}
y = \left\{ \begin{aligned}
&{\bf{w}}_{\rm{R}}^{\rm{H}}{\bf{n}}\thicksim \mathcal{CN}(0,{N_0}),&{\cal{H}}_0\\
&\sqrt P {\bf{w}}_{\rm{R}}^{\rm{H}}{\bf{H}}_0{{\bf{w}}_{\rm{T}}}s + {\bf{w}}_{\rm{R}}^{\rm{H}}{\bf{n}}\thicksim \mathcal{CN}(S,{N_0}),&{\cal{H}}_1
\end{aligned} \right.
\end{equation}
where ${\cal{H}}_0$ and ${\cal{H}}_1$ represent the cases when the AoD does not locate and locates within $[\psi_0,\psi_0+B]$, respectively, $S=\sqrt P {\bf{w}}_{\rm{R}}^{\rm{H}}{\bf{H}}_0{{\bf{w}}_{\rm{T}}}s$ denotes the received pure signal. Given a threshold $\Gamma N_0$, the instantaneous DP is given by
\begin{equation}
\begin{aligned}
p_{\rm{D}}(\Gamma) &= \mathrm{Pr}\{|(y|{\cal{H}}_1)|^2>\Gamma N_0\}=\mathrm{Pr}\{|S+n|^2>\Gamma N_0\},
\end{aligned}
\end{equation}
where $n={\bf{w}}_{\rm{R}}^{\rm{H}}{\bf{n}}$. To derive the average DP, we need to average $p_{\rm{D}}(\Gamma)$ on all the random variables. Note that $S$ depends on ${\bf{H}}_0$ and ${\bf{H}}_0$ depends on $\lambda$, $\Omega$ and $\psi$. Hence, we need to average $p_{\rm{D}}(\Gamma)$ on $n$, $\lambda$, $\Omega$ and $\psi$.

Let us first fix $\Omega$ and $\psi$ and average $p_{\rm{D}}(\Gamma)$ on $n$ and $\lambda$. Since $S=\sqrt P {\bf{w}}_{\rm{R}}^{\rm{H}}{\bf{H}}_0{{\bf{w}}_{\rm{T}}}s$, when $\Omega$ and $\psi$ are fixed, ${\bf{H}}_0$ has only one random parameter $\lambda\thicksim \mathcal{CN}(0,1)$. In such a case, $S$ can be seen as a zero-mean complex Gaussian variable, and $(S+n)\thicksim \mathcal{CN}(0,(1+\gamma){N_0})$, where $\gamma$ denotes the average received SNR given by
\begin{equation}
\begin{aligned}
\gamma &=\mathbb{E}_\lambda\left\{|\sqrt P {\bf{w}}_{\rm{R}}^{\rm{H}}{\bf{H}}_0{{\bf{w}}_{\rm{T}}}s|^2/N_0 \right\}\\
&= \frac{P {{{M_{\rm{AN}}}}{{N_{\rm{AN}}}}}}{N_0}|{\bf{w}}_{\rm{R}}^{\rm{H}}{\bf{a}}(N_{\rm{AN}},\Omega){\bf{a}}(M_{\rm{AN}},\psi)^{\rm{H}}{{\bf{w}}_{\rm{T}}}|^2\\
&
=\frac{P}{N_0}|A({\bf{w}}_{\rm{T}},\psi)|^2|A({\bf{w}}_{\rm{R}},\Omega)|^2,
\end{aligned}
\end{equation}
where $|A({\bf{w}}_{\rm{T}},\psi)|$ and $|A({\bf{w}}_{\rm{R}},\Omega)|$ are in fact Tx and Rx array gains depending on $\psi$ and $\Omega$, respectively. According to \cite[Chapter 2]{proakisdigital}, $|S+n|^2/N_0$ obeys Chi-square distribution with 2 degrees, and its cumulative distribution function (CDF) is $F(y)=1-e^{-y/(1+\gamma)}$. Thus, we have
\begin{equation} \label{eq_PD_lambdanoise}
\bar{p}_{\rm{D0}}(\Gamma) = 1-F(\Gamma) = e^{-\Gamma/(1+\gamma)}.
\end{equation}

\eqref{eq_PD_lambdanoise} is the result of averaging DP on $\lambda$ and $n$. We need to further average $\bar{p}_{\rm{D0}}(\Gamma)$ in \eqref{eq_PD_lambdanoise} on $\Omega$ and $\psi$ to obtain the ultimate average DP. Note that as we only want to evaluate the quality of the Tx codeword ${\bf{w}}_{\rm{T}}$ with angle coverage $[\psi_0,\psi_0+B]$, we can first get rid of the effects of the Rx codeword ${\bf{w}}_{\rm{R}}$ and AoA $\Omega$. Consequently, we assume the RX gain is fixed for simplicity, and without loss of generality we let $|A({\bf{w}}_{\rm{R}},\Omega)|^2=1$. Although the assumption may lead to inaccuracy, it simplifies the ultimate GDP metric, which does not require an accurate average DP expression. As a result, $\gamma$ reduces to
\begin{equation}
\gamma = \frac{P}{N_0}|A({\bf{w}}_{\rm{T}},\psi)|^2.
\end{equation}
And considering the MTP of ${\bf{w}}_{\rm{T}}$, the maximal received SNR is
\begin{equation}
\begin{aligned}
\gamma_{\rm{MAX}} &= \frac{P_{\rm{MAX}}}{N_0}|A({\bf{w}}_{\rm{T}},\psi)|^2\\
&=\frac{P_{\rm{PER}}}{{\|{{\bf{w}}_{\rm{T}}}\|_\infty^2}N_0}|A({\bf{w}}_{\rm{T}},\psi)|^2\\
&\triangleq \frac{\gamma_{\rm{PER}}}{{\|{{\bf{w}}_{\rm{T}}}\|_\infty^2}}|A({\bf{w}}_{\rm{T}},\psi)|^2,
\end{aligned}
\end{equation}
where $\gamma_{\rm{PER}}$ denotes the per-antenna received SNR under the PAPC.

Consequently, the average DP is given by
\begin{equation}\label{eq_metric_DP}
\begin{aligned}
&\bar{p}_{\rm{D}}(\Gamma)= \frac{1}{B} \int_{\psi_0}^{\psi_0+B} e^{-\Gamma/(1+\gamma_{\rm{MAX}})}d\psi\\
=&\frac{1}{B} \int_{\psi_0}^{\psi_0+B} \exp \left(-\frac{\Gamma}{1+\frac{\gamma_{\rm{PER}}}{{\|{{\bf{w}}_{\rm{T}}}\|_\infty^2}}|A({\bf{w}},\psi)|^2} \right)d\psi.
\end{aligned}
\end{equation}

We can see that $\bar{p}_{\rm{D}}(\Gamma)$ depends on both $\Gamma$ and $\gamma_{\rm{PER}}$ in addition to the codeword ${\bf{w}}_{\rm{T}}$ itself. Hence, it cannot be directly used as a general metric to evaluate the quality of a codeword. However, we can define one based on $\bar{p}_{\rm{D}}(\Gamma)$. Firstly, the threshold $\Gamma$ affects only the tradeoff between DP in hypothesis ${\cal{H}}_1$ and false-alarm probability (FAP) in hypothesis ${\cal{H}}_0$ \cite{xiao2013glrt}. When $\Gamma$ is smaller, DP is higher, but meanwhile FAP is also higher. In fact, the threshold itself does not affect the detection capability which involves both DA and FAP \cite{xiao2013glrt}. Based on this fact, we can just set $\Gamma=1$ without loss of generality. When $\Gamma$ is larger/smaller, DP will be lower/higher, but the comparison result of average DP between two different codewords basically maintains.

On the other hand, $\gamma_{\rm{PER}}$ may affect the comparison result of average DP between two different codewords. Intuitively, when $\gamma_{\rm{PER}}$ is sufficiently high, the beam pattern (reflected by $|A({\bf{w}},\psi)|$ in \eqref{eq_metric_DP}) is dominant, but when $\gamma_{\rm{PER}}$ is small, the maximal received SNR (reflected by ${\|{{\bf{w}}_{\rm{T}}}\|_\infty^2}$ in \eqref{eq_metric_DP}) is dominant. Hence, different system should set different $\gamma_{\rm{PER}}$.

A possible way is to set a typical value of $\gamma_{\rm{PER}}$ based on the system settings. For instance, the saturation power of a PA can be set to 15 dBm \cite{jin2007millimeter,zhao2012wideband}. According to the Friis formula, when the wavelength of the carrier frequency is 1 centimeter (30 GHz), and the Tx/Rx distance is 100 meters, $P_{\rm{PER}}$ will be $15-20\log_{10}(4\pi\times 10000)=-87$ dBm. Besides, when the bandwidth $B=100$ MHz, the noise power can be computed as $N_0=10\log_{10}(\kappa TB)=10\log_{10}(1.38\times 10^{-23}\times 300\times 10^8\times 10^3)=-74$ dBm, where $\kappa$ and $T$ are the Boltzmann constant and ambient temperature, respectively. Hence, the per-antenna received SNR is $(-76)-(-87)=-11$ dB. However, in the computation the spreading gain of the training sequence, that equals to the length of the training sequence, is not taken into account. If the length of the training sequence is $L_{\rm{S}}=128$, the spreading gain is $10\log_{10}{128}=21$ dB. If the possible propagation loss due to reflection, blockage etc. is 0 to 15 dB, $\gamma_{\rm{PER}}$ will have a dynamic range from $(21-11-15)$ dB to $(21-11-0)$ dB, i.e., $[-5,10]$ dB, according to the above evaluation. In this paper we prefer to set $\gamma_{\rm{PER}}=0$ dB for conciseness, but it should be clarified that other typical values close to 0 dB are also applicable. It will be shown in Section VI that a small change of $\gamma_{\rm{PER}}$ does not affect the comparison result of two codewords. For mmWave PA that provides a higher/lower saturation power, $\gamma_{\rm{PER}}$ can be set to $(\eta-15)$ dBm, where $\eta$ is the saturation power with unit dBm.

Based on the above discussions, we propose the metric of generalized detection probability (GDP) for an arbitrary $N$-entry Tx codeword ${\bf{w}}$ with unit 2-norm and target coverage $[\psi_0,\psi_0+B]$:
\begin{equation} \label{eq_Metric}
\begin{aligned}
&\xi ({\bf{w}},\psi_0,B)= \frac{1}{B} \int_{\psi_0}^{\psi_0+B} \exp \left(-\frac{{\|{{\bf{w}}}\|_\infty^2}}{{\|{{\bf{w}}}\|_\infty^2}+{|A({\bf{w}},\psi)|^2}} \right)d\psi,
\end{aligned}
\end{equation}
where ${\|{{\bf{w}}}\|_\infty^2}={\max(\{|[{\bf{w}}]_n|^2\}_{n=1}^{N})}$.

Note that although \eqref{eq_Metric} is defined for Tx codewords, it also can be used for Rx codewords, because small input fluctuation of low-noise amplifier (LNA) is also favored in mmWave communications, where the linearity of LNA may be not perfect due to the high frequency and large signal bandwidth. In the case that the linearity of LNA is good enough, \eqref{eq_Metric} can be modified by replacing ${\|{{\bf{w}}}\|_\infty^2}$ with constant 1 for Rx codewords. In this paper, we use \eqref{eq_Metric} for both Tx/Rx codeword designs.

\subsection{Properties and Significance}

The GDP metric has the following properties.

\textbf{Property 1: (Phase-shift invariance)} $\xi ({\bf{w}},\psi_0,B)$ is invariant to phase shift, i.e., $\xi ({\bf{w}}\circ \sqrt N {\bf{a}}(N,\Omega ),\psi_0+\Omega,B)=\xi ({\bf{w}},\psi_0,B)$, where $\circ$ represents entry-wise product (a.k.a. Hadamard product) and all angles are in the cosine angle domain.
\begin{proof}
See Appendix A.
\end{proof}

\textbf{Property 2: (In favor of CA weights)} $\xi ({\bf{w}},\psi_0,B)$ increases with ${\|{{\bf{w}}}\|_\infty^2}$ decreases, and $\xi ({\bf{w}},\psi_0,B)\leq \frac{1}{B} \int_{\psi_0}^{\psi_0+B} \exp \left(-\frac{1}{1+N{|A({\bf{w}},\psi)|^2}} \right)d\psi$, where the equality holds only when $|[{\bf{w}}]_n|^2=1/N$.
\begin{proof}
It is clear that $\exp \left(-\frac{{\|{{\bf{w}}}\|_\infty^2}}{{\|{{\bf{w}}}\|_\infty^2}+{|A({\bf{w}},\psi)|^2}} \right)$ increases with ${\|{{\bf{w}}}\|_\infty^2}$ decreases. Thus $\xi ({\bf{w}},\psi_0,B)$ increases with ${\|{{\bf{w}}}\|_\infty^2}$ decreases. In addition,
\[
\begin{aligned}
{\|{{\bf{w}}}\|_\infty^2}&={\max(\{|[{\bf{w}}]_n|^2\}_{n=1}^{N})}\geq  {{\rm{mean}}(\{|[{\bf{w}}]_n|^2\}_{n=1}^{N})}=1/N,
\end{aligned}
\]
where the equality holds only when $|[{\bf{w}}]_n|^2=1/N$.
\end{proof}

\textbf{Property 3: (In favor of flat beam pattern)} $\xi ({\bf{w}},\psi_0,B)\leq \exp \left(-\frac{{\|{{\bf{w}}}\|_\infty^2}}{{\|{{\bf{w}}}\|_\infty^2}+2/B} \right)$, and the equality holds only when ${\bf{w}}$ has an ideal beam pattern, i.e.,
\begin{equation}
{|A({\bf{w}},\psi)|^2}=\left\{
\begin{aligned}
&2/B,~~\psi \in [\psi_0,\psi_0+B],\\
&0,~~~~~~{\rm{Others}}.
\end{aligned}\right.
\end{equation}

\begin{proof}
See Appendix B.
\end{proof}


These properties of GDP offer guidance on codebook design. Property 1 shows that if a codeword ${\bf{w}}$ has been designed with coverage $[\psi_0,\psi_0+B]$, another codeword with target coverage $[\psi,\psi+B]$ can be immediately obtained as ${\bf{w}}\circ \sqrt N {\bf{a}}(N,\psi-\psi_0)$ without re-launching the designing process.
Property 2 implies that a good codeword should have elements with close amplitudes, such that the MTP will be higher. Property 3 shows that a good codeword should have equivalent beam gains along different angles (i.e., flat beam pattern); thus deep sinks with the beam pattern should be avoided.

Moreover, one significance of GDP lies in that it enables a general optimization approach to design the codewords. In particular, if we want to design an arbitrary codeword ${\bf{w}}$ with target beam coverage $[\psi_0,\psi_0+B]$, we can formulate the following problem
\begin{equation}\label{eq_optimization_general}
\begin{aligned}
\mathop {{\rm{maximize}}}\limits_{{\textbf{b}}}~~~&\xi ({\bf{w}}({\textbf{b}}),\psi_0,B),\\
{\rm{subject~to}}~~~&{\rm{Constraints~on~}}({\textbf{b}}),
\end{aligned}
\end{equation}
where ${\textbf{b}}$ is a parameter vector to be determined, and the constraints can include other desired structure constraints to simplify the search complexity in addition to the CA constraint. The proposed BMW-MS/LCS codebook is just obtained with the optimization approach (cf. \eqref{eq_optimization}). Note that the GDP metric and the optimization approach can be used for codebook design with both analog beamforming and hybrid precoding structures.

Another significance of GDP lies in that it provides an additional way to compare two different codewords/codebooks besides simulation. With the same target beam coverage, a codeword with higher GDP has better performance. For two different codebooks with the same coverage structure, its performance is basically determined by the codewords with the widest beams. Hence, by comparing the GDAs of the widest codewords of two different codebooks, we can evaluate which one is better. In Section VI we show that the results of GDP comparison agree with those of the success rate and achievable rate comparisons.

\section{Hierarchical Codebook Design}


In this section we propose the BMW-MS approach to design a Tx hierarchical codebook based on multi-RF-chain sub-array technique\footnote{Rx codebook design is similar.}. In order to obtain the coefficients for each sub-array, we propose two candidate solutions. The first one is a low-complexity search (LCS) solution to optimize the GDP metric, and the second one is a closed-form (CF) solution which is based on Property 3 to pursue flat beam patterns. Hence, the BMW-MS approach with the two solutions are termed as BMW-MS/LCS and BMW-MS/CF, respectively.

It is noteworthy that when letting $|{{\bf{f}}_{\rm{BB}}}|={\bf{1}}$ in \eqref{eq_codeword_Tx} the structure of the Tx codeword can be further written as
\begin{equation} \label{eq_codeword_structure}
{\bf{w}}={\sum_{i = 1}^{{M_{\rm{RF}}}} {{{\bf{v}}_i}} },
\end{equation}
where ${{\bf{v}}_i}=[{{\bf{F}}_{\rm{RF}}}]_{:,i}$ is the RF weight vector (RWV) of the $i$-th RF chain, and the phases of ${{\bf{f}}_{\rm{BB}}}$ have been absorbed into those of ${{{\bf{v}}_i}}$; thus we have in fact let ${{\bf{f}}_{\rm{BB}}}={\bf{1}}$ here.

\subsection{The BMW-MS Approach}

A critical challenge to design the hierarchical codebook shown in Fig. \ref{fig:codebook} is beam widening, i.e., to design the low-layer codewords which have wide beam widths.
Intuitively, if $M_{\rm{RF}}$ is sufficiently large, wide beams can be shaped by steering these RF RWVs towards equally spaced angles within the beam coverage. This is just the PS-DFT approach \cite{song2015multiRes}. However, in practice $M_{\rm{RF}}$ may be rather small. In such a case, we consider to use the sub-array technique to shape a wide beam. In particular, a large RWV of each RF chain can be divided into multiple sub-vectors (called sub-arrays), and these sub-arrays can point at different directions, such that a wider beam can be shaped.

To illustrate this, let us separate the $N$-element RWV of each RF chain into $M_{\rm{S}}$ sub-arrays with $N_{\rm{S}}$ elements in each sub-array, which means $N=M_{\rm{S}}N_{\rm{S}}$. In addition, letting ${{\bf{f}}_{i,m}} = {[{{\bf{v}}_i}]_{(m - 1){N_{\rm{S}}} + 1:m{N_{\rm{S}}}}}$, we have ${[{{\bf{f}}_{i,m}}]_n} = {[{{\bf{v}}_i}]_{(m - 1){N_{\rm{S}}} + n}}$, $m=1,2,...,M_{\rm{S}}$, $n=1,2,...,N_{\rm{S}}$, and $i=1,2,...,M_{\rm{RF}}$. ${{\bf{f}}_{i,m}}$ can be seen as the sub-RWV of the $m$-th sub-array of the $i$-th RF chain. Therefore, the beam gain of ${\bf{w}}$ writes
\begin{equation} \label{eq_beam_gain}
\begin{aligned}
&A({\bf{w}},\omega ) = \sum\limits_{n = 1}^N {{{[\sum\limits_{i = 1}^{{M_{\rm{RF}}}} {{{\bf{v}}_i}} ]_n}}{e^{ - j\pi (n - 1)\omega }}} \\
=& \sum\limits_{n = 1}^N {\sum\limits_{i = 1}^{{M_{\rm{RF}}}} {[{{\bf{v}}_i}} {]_n}{e^{ - j\pi (n - 1)\omega }}} \\
=& \sum\limits_{m = 1}^{{M_{\rm{S}}}} {\sum\limits_{n = 1}^{{N_{\rm{S}}}} {\sum\limits_{i = 1}^{{M_{\rm{RF}}}} {[{{\bf{v}}_i}} {]_{(m - 1){N_{\rm{S}}} + n}}{e^{ - j\pi ((m - 1){N_{\rm{S}}} + n - 1)\omega }}} } \\
=& \sum\limits_{i = 1}^{{M_{\rm{RF}}}} \sum\limits_{m = 1}^{{M_{\rm{S}}}} {\sum\limits_{n = 1}^{{N_{\rm{S}}}} {{e^{ - j\pi (m - 1){N_{\rm{S}}}\omega }}{{[{{\bf{f}}_{i,m}}]}_n}{e^{ - j\pi (n - 1)\omega }}} } \\
=&\sum\limits_{i = 1}^{{M_{\rm{RF}}}} {\sum\limits_{m = 1}^{{M_{\rm{S}}}} {{e^{ - j\pi (m - 1){N_{\rm{S}}}\omega }}A({{\bf{f}}_{i,m}},\omega )} },
\end{aligned}
\end{equation}
where we can find that the beam coverage of ${\bf{w}}$ can be controlled by controlling the $M_{\rm{RF}}M_{\rm{S}}$ sub-arrays ${\bf{f}}_{i,m}$. It is noteworthy that the coefficient between different sub-arrays is $e^{ - j\pi (m - 1)N_{\rm{S}}\omega }$. As the coefficient depends on $m$ and $\omega$, it induces coupling effect between different sub-arrays of the same RF chain. When the angle gap of two adjacent sub-arrays of the same RF chain is not wide enough, the coupling effect will be significant. In contrast, the coefficient does not depend on $i$. Hence, there is no coupling effect between different sub-arrays of different RF chains, which means that the steering angles of two sub-arrays of different RF chains can be close without affecting each other.

Based on the above observation, we propose the BMW-MS approach for beam widening, i.e., to cover an arbitrary angle range $[\Omega_0,\Omega_0+B]$ with $M_{\rm{RF}}$ RF chains, where each RF chain is decomposed into $M_{\rm{S}}$ sub-arrays, and the sub-RWVs ${\bf{f}}_{i,m}$ are set to steer along the angles
\begin{equation} \label{eq_BMW-MS_anglesteering}
\omega_{i,m}=\Omega_0+(i-1/2)\Delta\theta+(m-1)M_{\rm{RF}}\Delta\theta,
 \end{equation}
where $\Delta\theta=B/(M_{\rm{RF}}M_{\rm{S}})$, i.e., ${\bf{f}}_{i,m}$ satisfies
\begin{equation} \label{eq_BMW-MS_subAWV_set}
{\bf{f}}_{i,m}=\sqrt{\frac{{N_{\rm{S}}}}{N}}e^{j\theta_{i,m}}{\bf{a}}(N_{\rm{S}},\omega_{i,m}),
 \end{equation}
where $\theta_{i,m}$ are phase parameters (in the angle domain instead of cosine angle domain) to be determined. Since the beam width of a sub-array is $2/N_{\rm{S}}$, $\Delta\theta$ should be no larger than $2/N_{\rm{S}}$; otherwise there will be sink between two adjacent sub-arrays.

\begin{figure}[t]
\begin{center}
  \includegraphics[width=\figwidth cm]{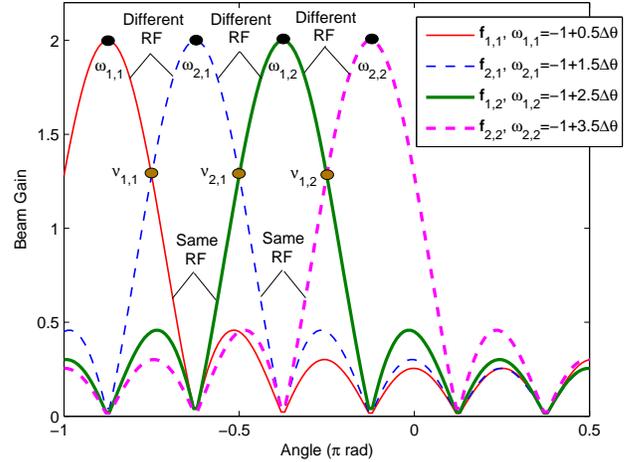}
  \caption{The beam patterns of the sub-arrays, where $N_{\rm{S}}=8$, $M_{\rm{RF}}=M_{\rm{S}}=2$, $\Delta\theta=2/N_{\rm{S}}=0.25$, $B=1$, and $\Omega_0=-1$.}
  \label{fig:subSteering}
\end{center}
\end{figure}

An example of the beam patterns of the sub-arrays is shown in Fig. \ref{fig:subSteering}, where $N_{\rm{S}}=8$, $M_{\rm{RF}}=M_{\rm{S}}=2$, $\Delta\theta=2/N_{\rm{S}}=0.25$, $B=1$, and $\Omega_0=-1$. The intuition of this approach is explained as follows. As we want to cover an angle interval of $B$, and there are $M_{\rm{RF}}M_{\rm{S}}$ controllable sub-arrays in total, we can evenly steer these sub-arrays with an angle gap $\Delta\theta=B/(M_{\rm{RF}}M_{\rm{S}})$ over the desired angle range. Moreover, in order to reduce the coupling effect between adjacent sub-arrays of the same RF chain, we set their angle gap as wide as possible.

\subsection{Low-Complexity Search and Closed-Form Solutions}

A remaining critical issue is to determine the coefficients $\theta_{i,m}$ for the BMW-MS approach in \eqref{eq_BMW-MS_subAWV_set}. We propose two solutions as follows, i.e., a low-complexity search (LCS) and a closed-form (CF) solutions.

\subsubsection{A Low-Complexity Search Solution}
According to \eqref{eq_optimization_general}, the following optimization problem can be formulated:
\begin{equation}\label{eq_optimization}
\begin{aligned}
\mathop {{\rm{maximize}}}\limits_{\theta_{i,m}}~~&\xi ({\bf{w}},\Omega_0,B),\\
{\rm{subject~to}}~~&{[{{\bf{v}}_i}]_{(m - 1){N_{\rm{S}}} + 1:m{N_{\rm{S}}}}} = {{\bf{f}}_{i,m}} = \\ &\sqrt{\frac{{N_{\rm{S}}}}{N}}e^{j\theta_{i,m}}{\bf{a}}(N_{\rm{S}},\omega_{i,m}),
\end{aligned}
\end{equation}
which is a non-convex problem. Although the exhaustive grid search can be directly adopted to search over the feasible domains of $\theta_{i,m}$, it has a high computational complexity which grows exponentially with the total number of the sub-arrays. To lower the search complexity, we assume equal-difference phase sequences for the sub-arrays of the same RF chain and the sub-arrays of different RF chains, respectively, i.e., we let
\begin{equation}
\theta_{i,m}=m\phi_1+i\phi_2,
\end{equation}
where $\phi_1\in [0,2\pi]$ is the phase difference of the phase sequence for the sub-arrays of the same RF chain, and $\phi_2\in[0,2\pi]$ is the phase difference of the phase sequence for the sub-arrays of different RF chains. With this assumption, the problem \eqref{eq_optimization} reduces to a 2-parameter search problem, which does not grow with the total number of sub-arrays. Thus, the computational complexity becomes affordable.


\subsubsection{A Closed-Form Solution}
According to \eqref{eq_BMW-MS_anglesteering}, the sub-arrays are set to steer along $\omega_{i,m}$ with an angle gap $\Delta\theta$. This can only guarantee that the beam gains along these directions are high. According to Property 3, we also hope that the beam gains along the other angles between adjacent $\omega_{i,m}$ are high, such that the beam pattern is flatter. Thus, we can design $\theta_{i,m}$ to maximize the beam gains along the middle angles of adjacent $\omega_{i,m}$, i.e.,
\begin{equation} \label{eq_BMW-MS_middle_angle}
\nu_{i,m}=\Omega_0+i\Delta\theta+(m-1)M_{\rm{RF}}\Delta\theta,
\end{equation}
where $im\neq M_{\rm{RF}}M_{\rm{S}}$. Fig. \ref{fig:subSteering} also shows the locations of $\nu_{i,m}$.

Since ${{\bf{f}}_{i,m}} = \sqrt{\frac{{N_{\rm{S}}}}{N}}{e^{j{\theta _{i,m}}}}{\bf{a}}\left( {{N_{\rm{S}}},{\omega _{i,m}}} \right)$, according to \eqref{eq_beam_gain} the beam gain of ${\bf{w}}$ along angles $\nu_{i,m}$ can be derived as
\begin{equation} \label{eq_subarray_dev1}
\begin{aligned}
A({\bf{w}},\nu_{k,n} )=&\frac{{N_{\rm{S}}}}{\sqrt{N}}\sum\limits_{i = 1}^{{M_{\rm{RF}}}} \sum\limits_{m = 1}^{{M_{\rm{S}}}} {{e^{ - j\pi (m - 1){N_{\rm{S}}}\nu_{k,n} }}{e^{j{\theta _{i,m}}}}}\\
&{\bf{a}}{{({N_{\rm{S}}},\nu_{k,n} )}^{\rm{H}}}
{\bf{a}}\left( {{N_{\rm{S}}},\omega_{i,m} } \right).
\end{aligned}
\end{equation}

It is clear that to determine $e^{\theta_{i,m}}$ by optimizing the absolute beam gain in \eqref{eq_subarray_dev1} is still complicated. However, since we want to obtain a low-complexity solution, we do not directly solve the optimization problem. Noticing that $|{\bf{a}}{({N_{\rm{S}}},{\omega _1})^{\rm{H}}}{\bf{a}}({N_{\rm{S}}},{\omega _2})|$ becomes smaller when $|\omega_1-\omega_2|$ becomes greater from 0 to $2/{N_{\rm{S}}}$, and can be neglected when $|\omega_1-\omega_2|>2/{N_{\rm{S}}}$. This means that the two sub-arrays with steering angles closest to $\nu_{k,n}$ have the most significant effects on the beam gain along $\nu_{k,n}$, while the sub-arrays with steering angles far from $\nu_{k,n}$ have a little effect on the beam gain along $\nu_{k,n}$, which can also be observed from Fig. \ref{fig:subSteering}. This motivates us to consider only the two close sub-arrays when optimizing the beam gain for simplicity. With this idea, we can finally obtain
\begin{equation} \label{eq_theta_im}
\begin{aligned}
{\theta _{i,m}} &= \pi m(m - 1){N_{\rm{S}}}{M_{\rm{RF}}}\Delta \theta /2 - \\
&~~~~~~~\pi (m{M_{\rm{RF}}} + i)({N_{\rm{S}}} - 1)\Delta \theta /2,
\end{aligned}
\end{equation}
where $\Delta \theta=B/({M_{\rm{RF}}}{M_{\rm{S}}})$, $i=1,2,...,M_{\rm{RF}}$, and $m=1,2,...,M_{\rm{S}}$. The detailed derivation can be found in Appendix B.

\subsection{Codebook Generation}
Up to now we have assumed that $M_{\rm{RF}}$, $M_{\rm{S}}$ and $N_{\rm{S}}$ are known in priori. However, in practice $M_{\rm{RF}}$ is given by the system setting, while $M_{\rm{S}}$ and $N_{\rm{S}}$ are in fact determined by the beam width $B$ of the codeword to be designed. In other words, $M_{\rm{S}}$ and $N_{\rm{S}}$ may be different for different codewords with different beam widths. Since when $M_{\rm{S}}$ is smaller $N_{\rm{S}}$ will be bigger and a higher beam gain can be provided, $M_{\rm{S}}$ should be as small as possible. As $\Delta \theta=B/({M_{\rm{RF}}}{M_{\rm{S}}})\leq 2/N_{\rm{S}}$ and $N={M_{\rm{S}}}{N_{\rm{S}}}$, we can obtain
\begin{equation} \label{eq_MS}
{M_{\rm{S}}} = \left\lceil {\sqrt {BN/(2{M_{\rm{RF}}})} } \right\rceil,
\end{equation}
where $\lceil \cdot \rceil$ is the ceiling operation.

Recall that we need to design $\underline{{\bf{w}}}(k,n)$ instead of just ${\bf{w}}(k,n)$ itself. By exploiting the BMW-MS approach we can design $\underline{{\bf{w}}}(k,n)\triangleq({\bf{F}}_{{\rm{RF}}(k,n)},{\bf{f}}_{{\rm{BB}}(k,n)}={\bf{1}})$. Recall again that different codewords within the same composite codeword share the same ${\bf{F}}_{\rm{RF}}$, i.e., the same $\{{\bf{v}}_i\}_{i=1,2,...,M_{\rm{RF}}}$. This can be satisfied by using Property 1 as follows. According to Property 1,
\begin{equation}
\begin{aligned}
{\bf{w}}(k,n)&={\bf{w}}(k,1)\circ \sqrt N {\bf{a}}(N,\frac{2(n-1)}{{M_{\rm{RF}}^k}})\\
&={\bf{F}}_{{\rm{RF}}(k,1)}{\bf{1}}\circ \sqrt N {\bf{a}}(N,\frac{2(n-1)}{{M_{\rm{RF}}^k}})\\
&={\bf{F}}_{{\rm{RF}}(k,1)}\sqrt N {\bf{a}}(N,\frac{2(n-1)}{{M_{\rm{RF}}^k}}),
\end{aligned}
\end{equation}
which means all the codewords within the same layer can share the same ${\bf{F}}_{\rm{RF}}$.

In summary, the codebook is generated as follows, where $k=1,...,\log_{M_{\rm{RF}}}({M_{\rm{AN}}})$, $N=M_{\rm{AN}}$.
\begin{itemize}
  \item {Split each RF chain into $M_{\rm{S}}=\left\lceil {\sqrt {B_kN/(2{M_{\rm{RF}}})} } \right\rceil$ sub-arrays, where $B_k=2/{M_{\rm{RF}}^k}$. Let the number of antennas of each sub-array be $N_{\rm{S}}=N/M_{\rm{S}}$.}
  \item {Compute $\underline{{\bf{w}}}(k,1)$ as $({\bf{F}}_{{\rm{RF}}(k,1)}=\{{\bf{v}}_i\}_{i=1,2,...,M_{\rm{RF}}},~{\bf{1}})$, where ${[{{\bf{v}}_i}]_{(m - 1){N_{\rm{S}}} + 1:m{N_{\rm{S}}}}} = \sqrt{\frac{{N_{\rm{S}}}}{N}}e^{j\theta_{i,m}}{\bf{a}}(N_{\rm{S}},\omega_{i,m})$. $\omega_{i,m}$ is computed as \eqref{eq_BMW-MS_anglesteering} ($\Omega_0=-1$, $B=2/2^k$). $\theta_{i,m}$ can either be computed by solving \eqref{eq_optimization} with the low-complexity search method (BMW-MS/LCS) or according to the closed-form expression \eqref{eq_theta_im} (BMW-MS/CF).}
  \item {Compute $\underline{{\bf{w}}}(k,n)$ ($n=2,3,...,M_{\rm{RF}}^k$) according to Property 1 as $({\bf{F}}_{{\rm{RF}}(k,1)},\sqrt N {\bf{a}}(N,\frac{2(n-1)}{{M_{\rm{RF}}^k}}))$.}
\end{itemize}


Interestingly, BMW-MS happens to be a generalization of the PS-DFT method in \cite{song2015multiRes}. With the PS-DFT method, a wide beam is constructed by exploiting multiple RF chains, which steer toward adjacent angles with a gap $2/N$, and the shaped beam width is $B=2M_{\rm{RF}}/N$. To lower the search complexity, the CA coefficients corresponding to the RWVs of all the RF chains are modeled as an equal-difference sequence in terms of the phase, and thus there is only one single parameter to determine. A drawback of the PS-DFT method lies in that the number of RF chains is proportional to the beam width. To generate very-wide beam, e.g., the codeword ${\bf{w}}(0,1)$, the number of RF chains would be too many for implementation.

In BMW-MS $M_{\rm{RF}}$ is assumed small. However, in the case that the number of available RF chains is large enough, e.g., equal to $N$, to shape a codeword with a beam width of $B$, we can determine the number of RF chains to generate this codeword as $M_{\rm{RF}}=BN/2$. Then according to \eqref{eq_MS}, we obtain $M_{\rm{S}}=1$, and thus $N_{\rm{S}}=N$. In such a case, BMW-MS/LCS becomes almost the same as PS-DFT except the objective function. Moreover, according to \eqref{eq_theta_im} for BMW-MS/CF, in the case of $M_{\rm{S}}=1$, $\{\theta_{i,m}\}_{i=1}^{M_{\rm{RF}}}$ is just an increasing sequence, which has a good accordance with \cite{song2015multiRes} and BMW-MS/LCS. Therefore, we say BMW-MS can be seen as a generalization of PS-DFT.

However, we emphasize that the main purpose of BMW-MS is to design a full hierarchical codebook shown in Fig. \ref{fig:codebook} with as less as possible RF chains, because in reality the number of available RF chains in a mmWave device would be small, e.g., typically only 2, 4 or 8. In fact, we recommend to select $M_{\rm{RF}}=2$ to realize BMW-MS, because fewer RF chains help to reduce the input/output fluctuation of the PAs, and with 2 RF chains BMW-MS can already achieve promising performance as we shall see from simulations later, but a larger number of RF chains can improve the efficiency of channel estimation as shown in Section III.

\section{Performance Evaluation}
In this section we evaluate the performance of BMW-MS. We will first show the beam patterns of BMW-MS and compare them with those of the alternatives. Afterwards, we will perform extensive performance comparisons between these candidates in terms of the GDP metric, success (detection) rate and achievable rate.

\begin{figure}[t]
\begin{center}
  \includegraphics[width=\figwidth cm]{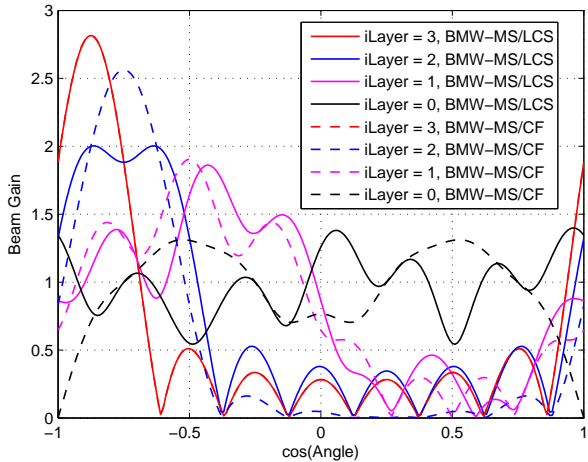}
  \caption{Beam pattern comparison between BMW-MS/LCS and BMW-MS/CF, where $N=8$, $M_{\rm{RF}}=2$, $iLayer$ is the layer index. The dashed red line superpositions the solid red line.}
  \label{fig:beamcomp}
\end{center}
\end{figure}

Fig. \ref{fig:beamcomp} shows the beam pattern comparison between BMW-MS/LCS and BMW-MS/CF, where $N=8$, $M_{\rm{RF}}=2$ and $iLayer$ is the layer index. From this figure we can observe that both approaches have realized the beam coverage shown in Fig. \ref{fig:codebook}. In addition, BMW-MS/CF basically provides similar beam patterns to BMW-MS/LCS, which means that the closed-form solution is also promising.

\begin{figure}[t]
\begin{center}
  \includegraphics[width=\figwidth cm]{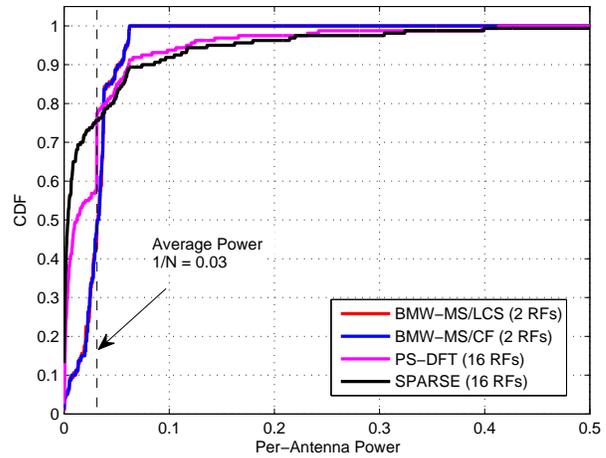}
  \caption{CDF comparison between different codebooks. $N=32$.}
  \label{fig:CDF}
\end{center}
\end{figure}

Next, we compare the performances of different codebooks. We know that when the saturation power of a PA is limited, the input fluctuation significantly affects the average output power. Basically the less the fluctuation is, the higher the average output power is. Hence, we first evaluate the fluctuation of the input power of the antennas with different codebooks. To do so, as each codebook has $\log_2(N)+1$ layers, we select one codeword from each layer of a codebook except the $0$th layer, because SPARSE did not provide the $0$th layer in \cite{alkhateeb2014channel}. Since each codeword has $N$ elements, we now have $N\log_2(N)$ elements in total. Hence, we can calculate the statistics on these elements and obtain the CDF curve. Fig. \ref{fig:CDF} shows the CDF comparison between different codebooks, where $N=32$. Note that in this figure all the codewords have unit 2-norm, i.e., the PAPC is not applied yet. We can find that for BMW-MS/LCS and BMW-MS/CF, most of the elements locate around the average power $1/N$, and the strongest element has a power about 0.06 (corresponding to ${\|{{\bf{w}}}\|_\infty^2}$ in \eqref{eq_MTP}). However, for PS-DFT and SPARSE, the power of the elements disperses within a large range from 0 to more than 0.5. Hence, it is clear that BMW-MS/LCS and BMW-MS/CF have lower power fluctuation than PS-DFT and SPARSE, and according to \eqref{eq_MTP} we can deduce that under the PAPC BMW-MS/LCS and BMW-MS/CF have higher MTP than PS-DFT and SPARSE.

\begin{figure}[t]
\begin{center}
  \includegraphics[width=\figwidth cm]{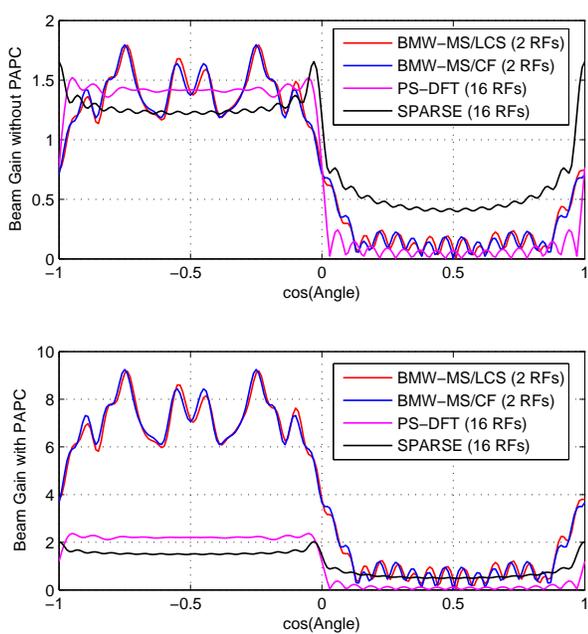}
  \caption{Beam pattern comparison between different schemes with/without PAPC. $N=32$, and the codeword is ${\bf{w}}(1,1)$ with coverage $[-1,0]$ for all the schemes.}
  \label{fig:beamcomp_diff}
\end{center}
\end{figure}

Fig. \ref{fig:beamcomp_diff} shows beam pattern comparison between different schemes with/without PAPC, where $N=32$ and the codeword is ${\bf{w}}(1,1)$ with coverage $[-1,0]$ for all the schemes. When without PAPC the 2-norm of a codeword is normalized to 1, while when with PAPC the entry with the largest absolute value of the codeword is normalized to 1 according to \eqref{eq_MTP}. From this figure we can find that without PAPC, PS-DFT and SPARSE have flatter beam patterns than BMW-MS/LCS and BMW-MS/CF (the upper figure), but with PAPC, BMW-MS/LCS and BMW-MS/CF have higher beam gains than PS-DFT and SPARSE (the bottom figure).

\begin{figure*}[t]
\begin{center}
  \includegraphics[width=14 cm]{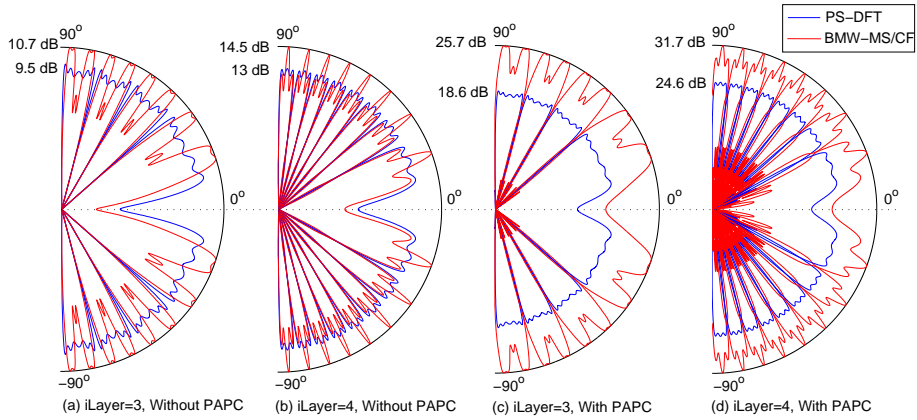}
  \caption{Beam pattern comparison between BMW-MS/CF and PS-DFT with/without PAPC. $N=64$. (a) and (b) are without PAPC, while (c) and (d) are with PAPC.}
  \label{fig:beamcomp_VS}
\end{center}
\end{figure*}

Fig. \ref{fig:beamcomp_VS} shows the beam pattern comparison between BMW-MS/CF and PS-DFT with/without PAPC, where $N=64$. (a) and (b) are without PAPC, and the beam patterns match the results in \cite[Fig.4 (b: Level 1) and (b: Level 2)]{song2015multiRes}. (c) and (d) are with PAPC. From this figure we can find that PS-DFT outperforms BMW-MS when there is no PAPC, because PS-DFT has flatter beam patterns. Note that PS-DFT achieves the superiority at the cost of a larger number of RF chains. The numbers of RF chains are $N/2^{iLayer}$ and 2 for PS-DFT and BMW-MS, respectively. In contrast, when PAPC is considered, which is practically reasonable in mmWave communications due to the limited performance of PA, BMW-MS/CF can offer an higher beam gain.

\begin{figure}[t]
\begin{center}
  \includegraphics[width=\figwidth cm]{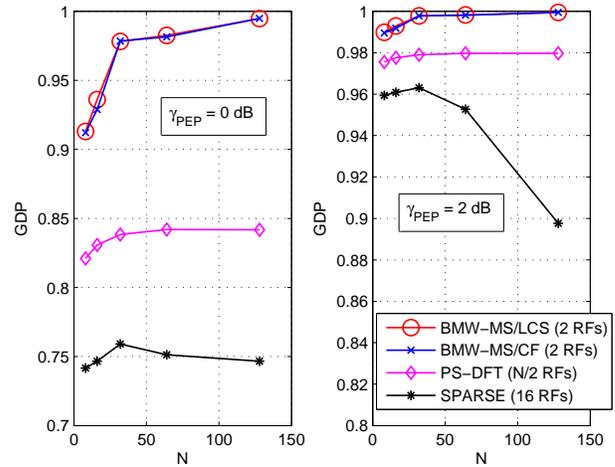}
  \caption{GDP comparison of the 1st layer codewords between different schemes under PAPC.}
  \label{fig:metricomp}
\end{center}
\end{figure}

Fig. \ref{fig:metricomp} shows the GDP comparison between different schemes under PAPC. The codewords of the 1st layer is considered, because the performance of a codebook is basically determined by the widest codeword\footnote{The 0th layer codeword was not provided in \cite{alkhateeb2014channel}. Hence we prefer to compare the 1st-layer codewords.}. Both cases of $\gamma_{\rm{PEP}} = 0$ dB (the left hand side figure) and $\gamma_{\rm{PEP}} = 2$ dB (the right hand side figure) are considered. From them we can observe that BMW-MS/LCS and BMW-MS/CF have equivalent GDP performance, which is significantly better than PS-DFT and SPARSE. For SPARSE, there is a peak GDP as $N$ increases. This is because when $N$ is small the received SNR plays a cardinal role to determine the GDP, and thus the GDP increases with $N$, which increases the beam gain. However, when $N$ is large, the received SNR is already high enough. In such a case, the beam pattern plays a cardinal role instead. Since the number of RF chains is fixed, there will appear sinks as $N$ increases according to \cite{alkhateeb2014channel}. Thus, the GDP decreases on the contrary as $N$ increases. Moreover, although the GDP grows when $\gamma_{\rm{PEP}} = 2$ dB compared with the case of $\gamma_{\rm{PEP}} = 0$ dB, the comparison results maintain, which demonstrates that the GDP metric in \eqref{eq_Metric} is reasonable and robust.

\begin{figure}[t]
\begin{center}
  \includegraphics[width=\figwidth cm]{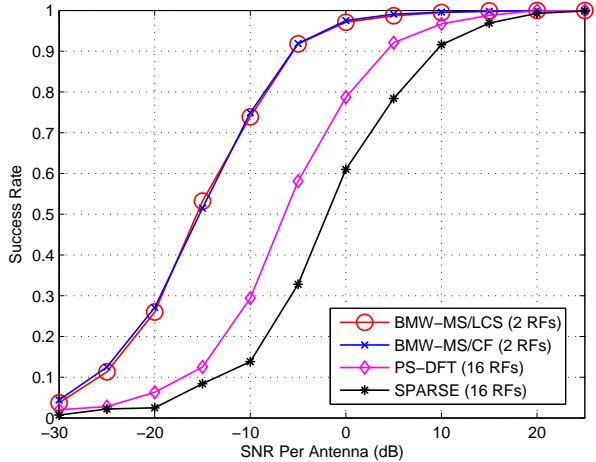}
  \caption{Comparison of success rate between different schemes under PAPC, where ${M_{\rm{AN}}}={N_{\rm{AN}}}=32$.}
  \label{fig:success_rate}
\end{center}
\end{figure}

\begin{figure}[t]
\begin{center}
  \includegraphics[width=\figwidth cm]{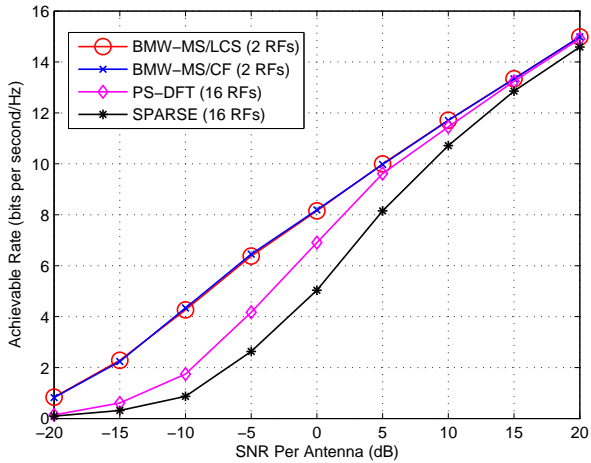}
  \caption{Comparison of achievable rate between different schemes under PAPC, where ${M_{\rm{AN}}}={N_{\rm{AN}}}=32$.}
  \label{fig:achievable_rate}
\end{center}
\end{figure}

Figs. \ref{fig:success_rate} and \ref{fig:achievable_rate} show the comparisons of success rate and achievable rate between different schemes under PAPC, where ${M_{\rm{AN}}}={N_{\rm{AN}}}=32$. $L=1$ in the simulations, and similar results can be observed when $L$ is set to other values. Success rate refers to the rate that an MPC is successfully acquired by Algorithm 1, while achievable rate is computed by using the best Tx/Rx precoding/combining codewords with Algorithm 1. These two figures show again that BMW-MS/LCS and BMW-MS/CF have similar overall performances, which are significantly better than PS-DFT and SPARSE. The results of these two figures have a good agreement with the GDP results shown in Fig. \ref{fig:metricomp}, which again demonstrates the rationality of the established GDP metric in \eqref{eq_Metric}.

\section{Conclusions}
In this paper we design a hierarchical codebook taking the per-antenna power constraint into account for mmWave channel estimation with a hybrid precoding/combining structure, where multiple RF chains are available. We have first established the GDP metric to particularly measure the performance of an arbitrary codeword. The metric not only enables a general optimization approach for codebook design, but also provides an additional way to compare the performance of different codewords/codebooks. Besides, GDP turns out to have a good accordance with the simulated performances of success (detection) rate and the achievable rate. To the best of our knowledge, GDP is the first metric that has been explicitly proposed for codebook design in mmWave communications. Then BWM-MS/LCS and BMW-MS/CF have been proposed to design a hierarchical codebook by exploiting the multi-RF-chain sub-array technique, where BWM-MS/LCS optimizes the GDP metric using a simplified search method under the sub-array structure, while BMW-MS/CF provides closed-form codewords to pursue flat beam patterns. Performance comparisons show that BMW-MS/LCS and BMW-MS/CF achieve very close performances, and they (with only 2 RF chains) outperform PS-DFT and SPARSE under the per-antenna power constraint.

\appendices
\section{Proof of Property 1}
Firstly, we have
\begin{equation} \label{eq_new_vector}
\begin{aligned}
&A({\bf{w}} \circ \sqrt N {\bf{a}}(N,\psi ),\Omega )\\
\mathop {=}\limits^{(a)}& \sqrt N {\bf{a}}{(N,\Omega )^{\rm{H}}}({\bf{w}} \circ \sqrt N {\bf{a}}(N,\psi ))\\
\mathop {=}\limits^{(b)}& \sum\limits_{n = 1}^N {{{[{\bf{w}}]}_n}{e^{j\pi (n - 1)\psi }}{e^{ - j\pi (n - 1)\Omega }}} \\
 =& \sum\limits_{n = 1}^N {{{[{\bf{w}}]}_n}{e^{ - j\pi (n - 1)(\Omega  - \psi )}}} \\
\mathop {=}\limits^{(c)}& A({\bf{w}},\Omega  - \psi ),
\end{aligned}
\end{equation}
where (a) and (c) are according to \eqref{eq_beam_gain_Define}, while (b) is according to definition of the entry-wise product. Besides, we have
\begin{equation}
\begin{aligned}
&{{\|({\bf{w}} \circ \sqrt N {\bf{a}}(N,\psi )\|_\infty^2}}\\
=&{\max(\{|[{\bf{w}} \circ \sqrt N {\bf{a}}(N,\psi )]_n|^2\}_{n=1}^{N})}\\
=&{\max(\{|[{\bf{w}}]_n e^{j\pi(n - 1)\psi}|^2\}_{n=1}^{N})}\\
=&{\max(\{|[{\bf{w}}]_n|^2\}_{n=1}^{N})}=\|{\bf{w}}\|_\infty^2.
\end{aligned}
\end{equation}
Hence,
\begin{equation}
\begin{aligned}
&\xi ({\bf{w}}\circ \sqrt N {\bf{a}}(N,\Omega ),\psi_0+\Omega,B)\\
=&\frac{1}{B} \int_{\psi_0+\Omega}^{\psi_0+\Omega+B} \exp \left(-\frac{{\|{{\bf{w}}}\|_\infty^2}}{{\|{{\bf{w}}}\|_\infty^2}+{|A({\bf{w}},\psi-\Omega)|^2}} \right)d\psi\\
=&\frac{1}{B} \int_{\psi_0}^{\psi_0+B} \exp \left(-\frac{{\|{{\bf{w}}}\|_\infty^2}}{{\|{{\bf{w}}}\|_\infty^2}+{|A({\bf{w}},\alpha)|^2}} \right)d\alpha\\
=&\xi ({\bf{w}},\psi_0+\Omega,B).
\end{aligned}
\end{equation}


\section{Proof of Property 3}

According to \eqref{eq_Metric}, ${\bf{w}}$ can be written as the following summation form:
\begin{equation} \label{eq_Metric_sum}
\begin{aligned}
\xi ({\bf{w}},\psi_0,B)&= \frac{1}{B} \sum_{i=1}^{N_{\rm{B}}} \exp \left(-\frac{C}{C+{|A({\bf{w}},\psi_i)|^2}} \right)\Delta\psi\\
&={\rm{mean}}\left( \exp \left(-\frac{C}{C+{|A({\bf{w}},\psi_i)|^2}} \right)\right),
\end{aligned}
\end{equation}
where $\Delta\psi$ is small, $C\triangleq{\|{{\bf{w}}}\|_\infty^2}$, $\psi_i=\psi_0+i\Delta\psi$, and ${N_{\rm{B}}}=B/\Delta\psi$.

Let $f(x)=\exp(-\frac{C}{C+x})$. Since $f'(x)|_{x\geq0}>0$, $f(x)$ is a concave function \cite{boyd2004convex}, and thus we have
\begin{equation}
\begin{aligned}
&{\rm{mean}}\left( \exp \left(-\frac{C}{C+{|A({\bf{w}},\psi_i)|^2}} \right)\right)\\
\leq&\exp \left(-\frac{C}{C+{\rm{mean}}\left({|A({\bf{w}},\psi_i)|^2}\right)} \right),
\end{aligned}
\end{equation}
where the equality holds only when ${|A({\bf{w}},\psi_i)|^2}=\mu$, i.e., a constant. In the following we will show that this constant is $\mu=2/B$.

Given an arbitrary codeword ${\bf{w}}$ with unit 2-norm, the average power of its beam gain in the angle domain is

\begin{equation}
\begin{aligned}
&{{\bar P}_A} = \frac{1}{2}\int_{ - 1}^1 {|A({\bf{w}},\Omega ){|^2}{\rm{d}}\Omega } \\
=& \frac{1}{2}\int_{ - 1}^1 {{{\left( {\sum\limits_{n = 1}^N {{{[{\bf{w}}]}_n}{e^{ - j\pi (n - 1)\Omega }}} } \right)}^{\rm{H}}}\left( {\sum\limits_{n = 1}^N {{{[{\bf{w}}]}_n}{e^{ - j\pi (n - 1)\Omega }}} } \right){\rm{d}}\Omega } \\
=& \frac{1}{2}\int_{ - 1}^1 {\sum\limits_{n = 1}^N {\sum\limits_{m = 1}^N {[{\bf{w}}]_n^{\rm{H}}{e^{j\pi (n - 1)\Omega }}{{[{\bf{w}}]}_m}{e^{ - j\pi (m - 1)\Omega }}} } {\rm{d}}\Omega } \\
=& \frac{1}{2}\int_{ - 1}^1 {\sum\limits_{n = 1}^N {\sum\limits_{m = 1}^N {[{\bf{w}}]_n^{\rm{H}}{{[{\bf{w}}]}_m}{e^{j\pi (n - m)\Omega }}} } {\rm{d}}\Omega } \\
=& \|{\bf{w}}{\|^2} + \frac{1}{2}\int_{ - 1}^1 {\sum\limits_{n = 1}^N {\sum\limits_{m = 1,m \ne n}^N {[{\bf{w}}]_n^{\rm{H}}{{[{\bf{w}}]}_m}{e^{j\pi (n - m)\Omega }}} } {\rm{d}}\Omega } \\
=& \|{\bf{w}}{\|^2} + \frac{1}{2}\sum\limits_{n = 1}^N {\sum\limits_{m = 1,m \ne n}^N {\int_{ - 1}^1 {[{\bf{w}}]_n^{\rm{H}}{{[{\bf{w}}]}_m}{e^{j\pi (n - m)\Omega }}{\rm{d}}\Omega } } } \\
\mathop  = \limits^{(a)}& \|{\bf{w}}{\|^2} + \frac{1}{2}\sum\limits_{n = 1}^N {\sum\limits_{m = 1,m \ne n}^N {[{\bf{w}}]_n^{\rm{H}}{{[{\bf{w}}]}_m}\int_{ - 1}^1 {{e^{j\pi (n - m)\Omega }}{\rm{d}}\Omega } } } \\
=& \|{\bf{w}}{\|^2}=1,
\end{aligned}
\end{equation}
where in (a) we have used
\begin{equation}
\int_{ - 1}^1 {{e^{j\pi (n - m)\Omega }}{\rm{d}}\Omega }  = \frac{{2j\sin (\pi (n - m))}}{{j\pi (n - m)}}\Big|_{m\neq n} = 0.
\end{equation}

Since $|A({\bf{w}},\psi)|^2=\mu$ when $\psi\in [\psi_0,\psi_0+B]$, and $|A({\bf{w}},\psi)|=0$ when $\psi \notin [\psi_0,\psi_0+B]$, we have $\mu B/2=1$. Hence $\mu=2/B$, which completes the proof.

\section{Derivation of \eqref{eq_theta_im}}
\begin{figure*}
\begin{equation} \label{eq_subarray_dev2}
\begin{aligned}
&A({\bf{w}},{\nu _{k,n}})|_{k < {M_{\rm{RF}}},~{{\bf{f}}_{i,m}} = {e^{j{\theta _{i,m}}}}{\bf{a}}\left( {{N_{\rm{S}}},{\omega _{i,m}}} \right)}=\frac{{N_{\rm{S}}}}{\sqrt{N}} \sum\limits_{i = 1}^{{M_{\rm{RF}}}} {\sum\limits_{m = 1}^{{M_{\rm{S}}}} {{e^{ - j\pi (m - 1){N_{\rm{S}}}{\nu _{k,n}}}}{e^{j{\theta _{i,m}}}}{\bf{a}}{{({N_{\rm{S}}},{\nu _{k,n}})}^{\rm{H}}}{\bf{a}}\left( {{N_{\rm{S}}},{\omega _{i,m}}} \right)} } \\
\approx& \frac{{N_{\rm{S}}}}{\sqrt{N}} {e^{ - j\pi (n - 1){N_{\rm{S}}}{\nu _{k,n}}}}{e^{j{\theta _{k,n}}}}{\bf{a}}{({N_{\rm{S}}},{\nu _{k,n}})^{\rm{H}}}{\bf{a}}\left( {{N_{\rm{S}}},{\omega _{k,n}}} \right) + \frac{{N_{\rm{S}}}}{\sqrt{N}} {e^{ - j\pi (n - 1){N_{\rm{S}}}{\nu _{k,n}}}}{e^{j{\theta _{k + 1,n}}}}{\bf{a}}{({N_{\rm{S}}},{\nu _{k,n}})^{\rm{H}}}{\bf{a}}\left( {{N_{\rm{S}}},{\omega _{k + 1,n}}} \right)\\
 =& \frac{1}{{\sqrt {{N}} }}{e^{ - j\pi (n - 1){N_{\rm{S}}}{\nu _{k,n}}}}{e^{j{\theta _{k,n}}}}{e^{ - j\pi ({N_{\rm{S}}} - 1)\Delta \theta /4}}\frac{{\sin ( - \pi \Delta \theta {N_{\rm{S}}}/4)}}{{\sin ( - \pi \Delta \theta /4)}} + \frac{1}{{\sqrt {{N}} }}{e^{ - j\pi (n - 1){N_{\rm{S}}}{\nu _{k,n}}}}{e^{j{\theta _{k + 1,n}}}}{e^{j\pi ({N_{\rm{S}}} - 1)\Delta \theta /4}}\frac{{\sin (\pi \Delta \theta {N_{\rm{S}}}/4)}}{{\sin (\pi \Delta \theta /4)}}\\
 =& \frac{1}{{\sqrt {{N}} }}\frac{{\sin (\pi \Delta \theta {N_{\rm{S}}}/4)}}{{\sin (\pi \Delta \theta /4)}}{e^{ - j\pi (n - 1){N_{\rm{S}}}{\nu _{k,n}}}}{e^{j{\theta _{k,n}}}}{e^{ - j\pi ({N_{\rm{S}}} - 1)\Delta \theta /4}}\left( {1 + {e^{j({\theta _{k + 1,n}} - {\theta _{k,n}} + \pi ({N_{\rm{S}}} - 1)\Delta \theta/2 )}}} \right).
\end{aligned}
\end{equation}
\hrulefill
\end{figure*}

\begin{figure*}
\begin{equation} \label{eq_subarray_dev3}
\begin{aligned}
&A({\bf{w}},{\nu _{{M_{\rm{RF}}},n}})|_{{{\bf{f}}_{i,m}} = {e^{j{\theta _{i,m}}}}{\bf{a}}\left( {{N_{\rm{S}}},{\omega _{i,m}}} \right)}=\frac{{N_{\rm{S}}}}{\sqrt{N}} \sum\limits_{i = 1}^{{M_{\rm{RF}}}} {\sum\limits_{m = 1}^{{M_{\rm{S}}}} {{e^{ - j\pi (m - 1){N_{\rm{S}}}{\nu _{k,n}}}}{e^{j{\theta _{i,m}}}}{\bf{a}}{{({N_{\rm{S}}},{\nu _{k,n}})}^{\rm{H}}}{\bf{a}}\left( {{N_{\rm{S}}},{\omega _{i,m}}} \right)} } \\
 \approx& \frac{{N_{\rm{S}}}}{\sqrt{N}} {e^{ - j\pi (n - 1){N_{\rm{S}}}{\nu _{k,n}}}}{e^{j{\theta _{k,n}}}}{\bf{a}}{({N_{\rm{S}}},{\nu _{k,n}})^{\rm{H}}}{\bf{a}}\left( {{N_{\rm{S}}},{\omega _{k,n}}} \right) + \frac{{N_{\rm{S}}}}{\sqrt{N}} {e^{ - j\pi n{N_{\rm{S}}}{\nu _{k,n}}}}{e^{j{\theta _{1,n + 1}}}}{\bf{a}}{({N_{\rm{S}}},{\nu _{k,n}})^{\rm{H}}}{\bf{a}}\left( {{N_{\rm{S}}},{\omega _{1,n + 1}}} \right)\\
 = &\frac{1}{{\sqrt {{N}} }}{e^{ - j\pi (n - 1){N_{\rm{S}}}{\nu _{k,n}}}}{e^{j{\theta _{k,n}}}}{e^{ - j\pi ({N_{\rm{S}}} - 1)\Delta \theta /4}}\frac{{\sin ( - \pi \Delta \theta {N_{\rm{S}}}/4)}}{{\sin ( - \pi \Delta \theta /4)}} + \frac{1}{{\sqrt {{N}} }}{e^{ - j\pi n{N_{\rm{S}}}{\nu _{k,n}}}}{e^{j{\theta _{1,n + 1}}}}{e^{j\pi ({N_{\rm{S}}} - 1)\Delta \theta /4}}\frac{{\sin (\pi \Delta \theta {N_{\rm{S}}}/4)}}{{\sin (\pi \Delta \theta /4)}}\\
 = &\frac{1}{{\sqrt {{N}} }}\frac{{\sin (\pi \Delta \theta {N_{\rm{S}}}/4)}}{{\sin (\pi \Delta \theta /4)}}{e^{ - j\pi (n - 1){N_{\rm{S}}}{\nu _{k,n}}}}{e^{j{\theta _{k,n}}}}{e^{ - j\pi ({N_{\rm{S}}} - 1)\Delta \theta /4}}\left( {1 + {e^{j[({\theta _{1,n + 1}} - {\theta _{k,n}} + \pi ({N_{\rm{S}}} - 1)\Delta \theta /2) - \pi {N_{\rm{S}}}{\nu _{k,n}}]}}} \right).
\end{aligned}
\end{equation}
\hrulefill
\end{figure*}

As shown in Fig. \ref{fig:subSteering}, there are two different types of positions of $\nu_{k,n}$. The first one is $\nu_{k,n}$ with $k=1,2,...,M_{\rm{RF}}-1$. The closest steering angles to it are $\omega_{k,n}$ and $\omega_{k+1,n}$, i.e., the two corresponding sub-arrays have adjacent RF indices and the same sub-array index. The other one is $\nu_{M_{\rm{RF}},n}$. The closest steering angles to it are $\omega_{M_{\rm{RF}},n}$ and $\omega_{1,n+1}$, i.e., the two corresponding sub-arrays have adjacent sub-array indices but the RF index switches from $M_{\rm{RF}}$ to 1. The beam gain of the first type of $\nu_{k,n}$ is derived as in \eqref{eq_subarray_dev2} on the top of the next page, where we have used
\begin{equation} \label{eq_sum_exp}
\begin{aligned}
&\sum\limits_{i = 1}^N {{e^{j(i - 1)\theta }}}  = \frac{{1 - {e^{jN\theta }}}}{{1 - {e^{j\theta }}}}=\frac{{{e^{jN\theta /2}}({e^{ - jN\theta /2}} - {e^{jN\theta /2}})}}{{{e^{j\theta /2}}({e^{ - j\theta /2}} - {e^{j\theta /2}})}}\\
 = &{e^{j(N - 1)\theta /2}}\frac{{\sin (N\theta /2)}}{{\sin (\theta /2)}}.
\end{aligned}
\end{equation}
From \eqref{eq_subarray_dev2} we can find that to optimize the absolute gain, we have
\begin{equation} \label{eq_theta_ralation2}
{\theta _{k + 1,n}} - {\theta _{k,n}} =  - \pi ({N_{\rm{S}}} - 1)\Delta \theta /2.
\end{equation}
In addition, the beam gain of the other type of $\nu_{k,n}$ is derived as in \eqref{eq_subarray_dev3} on the top of the next page, where we can find that to optimize the absolute gain, we have
\begin{equation} \label{eq_theta_ralation3}
\begin{aligned}
&{\theta _{1,n + 1}} - {\theta _{{M_{\rm{RF}}},n}}\\
 =&  - \pi ({N_{\rm{S}}} - 1)\Delta \theta /2 + \pi {N_{\rm{S}}}({M_{\rm{RF}}}\Delta \theta  + (n - 1){M_{\rm{RF}}}\Delta \theta )\\
 =&  - \pi ({N_{\rm{S}}} - 1)\Delta \theta /2 + \pi {N_{\rm{S}}}n{M_{\rm{RF}}}\Delta \theta
\end{aligned}
\end{equation}

Based on \eqref{eq_theta_ralation2} and \eqref{eq_theta_ralation3}, we finally obtain \eqref{eq_theta_im}.

%



\end{document}